\PassOptionsToPackage{dvipsnames}{xcolor}
\documentclass[sigconf, nonacm]{acmart}

\renewcommand\footnotetextcopyrightpermission[1]{} 

\usepackage{xcolor}
\usepackage[skip=3pt]{subcaption}
\usepackage[skip=3pt]{caption}
\usepackage{tabularx}
\usepackage{multirow}
\usepackage{pbalance}
\usepackage[a-2b,mathxmp]{pdfx}[2018/12/22]

\newcommand\vldbavailabilityurl{https://infosys.informatik.uni-mainz.de/rtindex}

\definecolor{darkyellow}{RGB}{255,168,67}
\definecolor{darkgreen}{RGB}{85,130,54}
\newcommand{\smalltt}[1]{\texttt{\small #1}}
\newcommand{\red}[1]{\textcolor{red}{#1}}
\newcommand{\yellow}[1]{\textcolor{darkyellow}{#1}}
\newcommand{\green}[1]{\textcolor{darkgreen}{#1}}

\definecolor{QueryBlue}{RGB}{68, 114, 196}
\definecolor{Triangle0}{RGB}{192, 0, 0}
\definecolor{Triangle1}{RGB}{234, 178, 0}
\definecolor{Triangle2}{RGB}{85, 130, 54}
\definecolor{Triangle3}{RGB}{112, 48, 160}
\definecolor{Triangle4}{RGB}{0, 77, 134}
\definecolor{Triangle5}{RGB}{0, 0, 0}
\definecolor{FloatSign}{RGB}{36, 90, 140}
\definecolor{FloatExponent}{RGB}{84, 130, 53}
\definecolor{FloatSignificand}{RGB}{177, 81, 15}
\definecolor{Qualifies}{RGB}{84, 130, 53}

\newcommand{\yes}{\textbf{\textcolor{darkgreen}{Y}}}
\newcommand{\no}{\textbf{\textcolor{Maroon}{N}}}

\newcolumntype{L}[1]{>{\raggedright\let\newline\\\arraybackslash\hspace{0pt}}m{#1}}
\newcolumntype{C}[1]{>{\centering\let\newline\\\arraybackslash\hspace{0pt}}m{#1}}
\newcolumntype{R}[1]{>{\raggedleft\let\newline\\\arraybackslash\hspace{0pt}}m{#1}}

\begin{document}
\title{RTIndeX: Exploiting Hardware-Accelerated GPU Raytracing for Database Indexing}

\author{Justus Henneberg}
\affiliation{%
  \institution{Johannes Gutenberg University}
  \city{Mainz}
  \country{Germany}}
\email{henneberg@uni-mainz.de}

\author{Felix Schuhknecht}
\affiliation{%
  \institution{Johannes Gutenberg University}
  \city{Mainz}
  \country{Germany}}
\email{schuhknecht@uni-mainz.de}

\begin{abstract}

Data management on GPUs has become increasingly relevant due to a tremendous rise in processing power and available GPU memory.
Similar to main-memory systems, there is a need for performant GPU-resident index structures to speed up query processing.
Unfortunately, mapping indexes efficiently to the highly parallel and hard-to-program hardware is challenging and often fails to yield the desired performance and flexibility. 
Instead of proposing yet another hand-tailored index, we investigate whether we can exploit an indexing mechanism that is already built into modern GPUs:
The raytracing hardware accelerator provided by NVIDIA RTX GPUs.
To do so, we re-phrase the database indexing problem as a raytracing problem, where we express the dataset to be indexed as objects in a 3D scene, and point/range lookups as rays across the scene.
In this combination, coined \textbf{RX} in the following, lookups are performed as intersection tests in hardware by dedicated raytracing cores.
To analyze the pros, cons, and usefulness of the raytracing pipeline for database indexing, we carefully evaluate \textbf{RX} along \textit{fourteen} dimensions and demonstrate its competitiveness and potential in a large variety of situations. 

\end{abstract}

\maketitle

\ifdefempty{\vldbavailabilityurl}{}{
\vspace{.3cm}
\begingroup\small\noindent\raggedright\textbf{Artifact availability:}\\
The source code, data, and/or other artifacts have been made available at \url{\vldbavailabilityurl}.
\endgroup
}

\thickmuskip=0.7\thickmuskip

\ifdefempty{\vldbavailabilityurl}{}{
\vspace{.3cm}
\begingroup\small\noindent\raggedright\textbf{Citing this research:}\\
This version is a preprint of the paper accepted for publication in PVLDB~Vol.~16. When citing this research, please cite the PVLDB version. 
\endgroup
}

\thickmuskip=0.7\thickmuskip

\section{Introduction}
\vspace*{-0.1cm}

Implementing performant index structures for highly-parallel GPU architectures is a challenging task~\cite{lit:rtree1, lit:rtree2, lit:radixtree, lit:fasttree, lit:lsmtree, lit:bitmapindex, lit:permutationindex, lit:learnedindex, lit:bloomfilter1, lit:bloomfilter2, lit:quotientfilter, lit:hash-dycuckoo, lit:hash-megakv, lit:btree1, lit:btree2, lit:hash-warpdrive, lit:hash-slabhash, lit:hash-cudpp1, lit:hash-cudpp2}.
But do we really have to implement a high-performing data structure from scratch?
Can we maybe utilize the hardware indexing mechanism that is \textit{already integrated} in modern GPUs?

\begin{figure*}[!t]
\centering

\begin{subtable}[b]{0.29\textwidth}
\centering
{
\small
\begin{tabular}{|c|c|c|}
\toprule
rowID & Article & Category  \\
\midrule
\textcolor{Triangle0}{0} & \textcolor{Triangle0}{Juice}   & \textcolor{Triangle0}{26 ($k_0$)}  \\
\textcolor{Triangle1}{1} & \textcolor{Triangle1}{Bread}   & \textcolor{Triangle1}{25 ($k_1$)}  \\
\textcolor{Triangle2}{2} & \textcolor{Triangle2}{Cookies} & \textcolor{Triangle2}{29 ($k_2$)}  \\
\textcolor{Triangle3}{3} & \textcolor{Triangle3}{Coffee}  & \textcolor{Triangle3}{23 ($k_3$)}  \\
\textcolor{Triangle4}{4} & \textcolor{Triangle4}{Donuts}  & \textcolor{Triangle4}{29 ($k_4$)}  \\
\textcolor{Triangle5}{5} & \textcolor{Triangle5}{Wine}    & \textcolor{Triangle5}{27 ($k_5$)}  \\
\bottomrule
\end{tabular}
}
\caption{Exemplary database table.}
\label{table:db}
\end{subtable}
\begin{subfigure}[b]{0.68\textwidth}
\includegraphics[page=1, width=1.0\linewidth, trim={1mm 148mm 190mm 4mm}, clip]{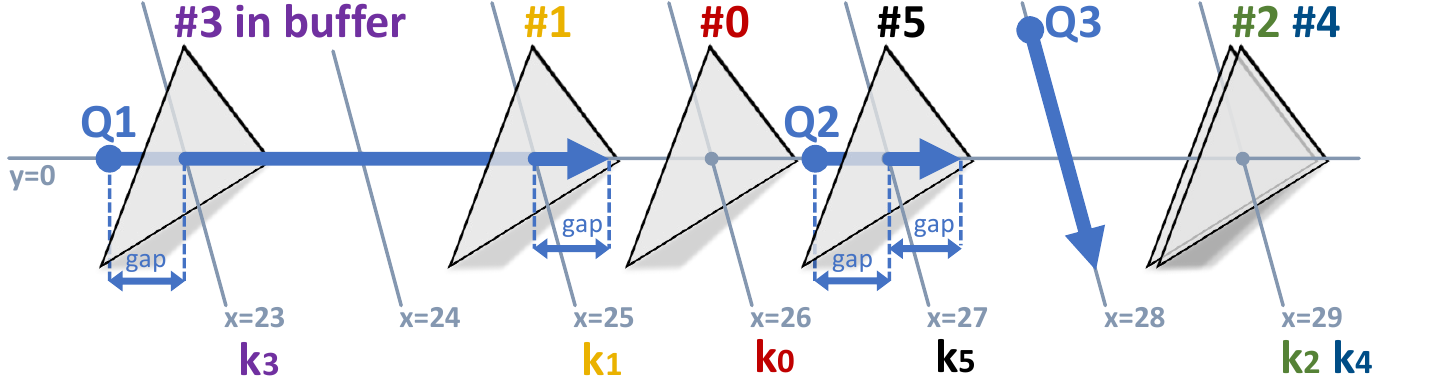} 
\caption{Corresponding triangle arrangement for the \textit{Category} column.}
\label{fig:core_principle}
\end{subfigure}
\caption{Visualization of how our indexing approach \textbf{RX} represents a secondary index on the \textit{Category} column. For each key~$k_i$ in \textit{Category}, we create a triangle centered around the point ($k_i$, 0, 0), where the triangles are stored internally in the same order as the keys they represent.
For each lookup, we fire a ray that is tested for intersection with all triangles. For example, the range lookup~Q1 tests for all keys in the range $[23,25]$ and consequently hits triangles \#3 and \#1, returning rowIDs 3 and 1.
}
\label{fig:intro}
\end{figure*}

\vspace*{-0.2cm}
\subsection{Hardware Accelerated Indexing on GPUs}
\vspace*{-0.1cm}

This indexing mechanism appears on NVIDIA's RTX workstation and consumer GPUs in form of a \textit{raytracing hardware accelerator}.
This accelerator enables rendering of ray-traced scenes in real time at a high frame rate.
The concept is simple:
A~3D~scene contains many objects, which are usually approximated by thousands of triangles each, and a virtual camera observing the scene from a certain position.
To create a realistic image, the GPU simulates the light rays entering the camera.
For efficiency reasons, rays are cast in reverse, i.e., they originate at the camera lens and then travel in the direction the camera is facing until they hit the closest object.
To speed up detecting intersections between rays and objects, graphics applications build a so-called \textit{bounding volume hierarchy (BVH)} over all objects of the scene. 
Using the BVH, modern GPUs can perform the intersection test efficiently in hardware for a large number of rays in parallel using specialized raytracing cores that exist solely for this purpose. 

Conceptually, finding intersections this way is nothing but a hardware-accelerated indexing mechanism. While the 3D objects in the scene resemble a dataset, the BVH serves as an auxiliary index structure on top of it. Casting a ray resembles a lookup:
If a ray intersects with an object, the lookup returns a unique identifier associated with the object.
This enables us to map other indexing problems, such as database indexing, to this mechanism, and to exploit the built-in hardware acceleration for fast lookups.
To create a secondary index on a table column, we express all entries in the column as 3D objects, ordered by their magnitude in the coordinate system of the scene.
We associate each object with the rowID of the corresponding entry in the table, which is retrieved when a ray hits the object.
To perform a lookup, we fire a ray through the area of interest, let the hardware detect all collisions, and return all associated rowIDs.
   
Unfortunately, expressing database indexing using this raytracing mechanism is not as straight-forward as it sounds at first glance and comes with a surprising amount of design choices to make. 
First, to encode each dataset entry as a 3D object, we use the powerful but non-trivial OptiX~\cite{lit:optix-paper, lit:optix-homepage} computing API, which allows us to freely program parts of the raytracing pipeline.
As OptiX provides numerous options to set up the scene, for example in the types of primitives to use and the way intersection tests are performed, we can identify several drastically different ways to express database indexing. 
Second, the raytracing hardware imposes certain restrictions that we have to respect.
For example, OptiX only supports single-precision floating-point numbers, while we want to index up to 64-bit wide integer keys.
Consequently, we have to work around this problem, and again, there are multiple different ways to do so. 
Third, the raytracing mechanism is a proprietary implementation by NVIDIA, where details about the internal structure and behavior are intentionally not made available to the public~\cite{lit:nvidia_forum}.
Therefore, it is highly unclear how well the problem of database indexing maps to the architecture and how it reacts to certain workloads, like dense/sparse key sets or the hit/miss ratio of lookups.

\vspace*{-0.4cm}
\subsection{Contributions and Structure of the Paper}
\vspace*{-0.1cm}

As a consequence of these observations, in the following work, we investigate whether and in which form hardware accelerated indexing can be used to realize database indexing on GPUs. 
We organize our work along \textit{fourteen} different dimensions composed of \textit{five} configuration dimensions and \textit{nine} experimental dimensions:

First, we present how to re-phrase database indexing as a raytracing problem.
Based on that, we discuss our implementation, coined \textbf{RX}, using the OptiX computing API.
\textbf{RX} supports hardware-accelerated point and range lookups on 64-bit integer columns on NVIDIA RTX GPUs.

Second, we discuss the configuration options of \textbf{RX} along five dimensions: We implement (1)~three different ways to express keys, (2)~three different types of scene primitives to express the indexed dataset, (3)~three different ways to express point and range lookups, (4)~a flexible key decomposition, and (5)~two different options to perform updates. Empirically, we identify the most suitable configuration that we will use throughout the rest of the paper.

Third, we perform an in-depth experimental evaluation where we compare \textbf{RX} against three GPU-resident index structures along nine experimental dimensions:
We vary (6)~the number of indexed keys and fired lookups, (7)~the multiplicity of the indexed keys, (8)~the order of the keys and lookups, (9)~the batch size, (10)~the hit/miss ratio, (11)~the selectivity of range lookups, (12)~the key size, and (13)~the distribution of keys and lookups.
Also, we (14)~compare the performance on the three latest hardware architectures.
Note that we perform a stand-alone evaluation of \textbf{RX} and its baselines to clearly identify the impact of the individual dimensions in isolation and without the interference of other system components.

\vspace*{-0.4cm}
\section{Database Indexing $\rightarrow$ Raytracing}
\label{sec:rtindex}
\vspace*{-0.1cm}

We start by re-phrasing database indexing as a raytracing problem. 
Figure~\ref{fig:intro} visualizes the high-level principle of the approach with a simple example:  
Assume we want to create a secondary index for the integer column \textit{Category} of the exemplary database table shown in Figure~\ref{table:db}.
To do so, we want to represent each key in the column by a corresponding primitive and associate it with its rowID.
For simplicity, we limit the discussion to triangles for now and discuss other primitive types in Section~\ref{ssec:primitives}.

\vspace*{-0.2cm}
\subsection{Building the Index}
\label{ssec:build}
\vspace*{-0.1cm}

\begin{figure}[h!]
\centering
\includegraphics[page=1, width=.85\linewidth, trim={230mm 0mm 0mm 148mm}, clip]{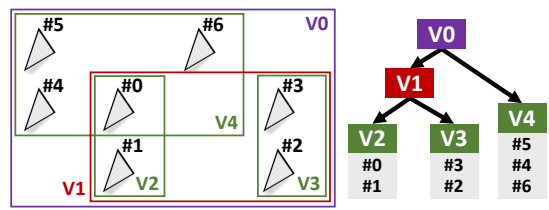}
\caption{Exemplary bounding volume hierarchy for a (two-dimensional) triangle arrangement.}
\label{fig:bvh}
\vspace*{-0.3cm}
\end{figure}

To build the index, we first convert each key $k_i$ of the table column into a corresponding triangle~$T_i$, where the 3D point $(k_i, 0, 0)$ should be a part of $T_i$.
In other words, we are using the numerical \textit{Category} value as our $x$-coordinate.
In the 3D scene, this results in a line of triangles with gaps of varying sizes in between, as visualized in Figure~\ref{fig:core_principle}.
If an entry occurs multiple times in the table (as is the case for key~$29$), multiple triangles will be created at the same location.
One way of constructing $T_i$ is by slightly offsetting each of the three triangle corners in a different direction, e.g., $(k_i, -0.5, -0.5)$, $(k_i+0.5, -0.5, 0.5)$, and $(k_i-0.5, 0.5, 0.5)$.
OptiX requires all triangles to be stored in a so-called \textit{vertex buffer}.
The position at which triangle~$T_i$ is stored in the vertex buffer is not arbitrary, but must correspond to its rowID $i$, as this position serves as the unique identifier that is returned by OptiX if a collision with $T_i$ is detected.
Once the vertices are arranged in the buffer, we can pass it to \smalltt{optixAccelBuild()} to generate the BVH.
A BVH is a tree-like data structure in which the individual triangles form the leaves of the tree.
These triangles are then combined into small disjoint groups.
For each group, the BVH stores a three-dimensional cuboid, called a \textit{bounding volume}, which encloses all triangles in the group.
These bounding volumes are then iteratively grouped and enclosed in the same way until only one group remains, forming the root of the tree.
Figure~\ref{fig:bvh} visualizes a bounding volume hierarchy for seven triangles in two dimensions.
On modern RTX GPUs, both the traversal of the BVH as well as the intersection tests between the ray and the candidate triangles in the bounding volume are hardware-accelerated.

\vspace*{-0.1cm}
\subsection{Performing Lookups}
\label{ssec:querying}
\vspace*{-0.1cm}

Now that the BVH is prepared, let us see how we utilize OptiX to answer lookups, where we start with the more general range lookups.
Conceptually, to answer a range lookup on the \textit{Category} column, such as $[23, 25]$~(\textbf{Q1}), we have to cast a ray along the line of triangles, starting just before the $x$-coordinate~$23$ and ending right after the $x$-coordinate~$25$.
In Figure \ref{fig:intro}, the range lookup ray hits triangles \textcolor{Triangle1}{\#1} and \textcolor{Triangle3}{\#3}, implying that rows with rowID~\textcolor{Triangle1}{$1$}~and~\textcolor{Triangle3}{$3$} in the original table satisfy the range predicate.
To realize point lookups, we can either formulate single-key range lookups~(as shown in \textbf{Q2}), or cast a short ray perpendicularly to the line of triangles~(as shown in \textbf{Q3}).
In practice, we can answer a large number of range lookups $[l^{(i)}, u^{(i)}]$ concurrently to exploit the parallel nature of the hardware.
Here, $l^{(i)}$ and $u^{(i)}$ denote the inclusive lower bound and upper bound, respectively, of the $i$-th range lookup. 
We formulate our batch of range lookups by specifying a corresponding ray for each pair of bounds, where each ray consists of a three-dimensional \textit{origin point}~$o$ and a \textit{direction vector}~$d$.
A ray intersects a triangle $T$ if there exists a $t > 0$ such that the point~$p = o + t \cdot d$ is part of the triangle $T$, where $t$ is called the \textit{intersection parameter}.
Note that we can restrict the intersection range by providing two additional parameters, $t_{\text{min}}$ and $t_{\text{max}}$.
In this case, we will only detect intersections that also satisfy $t_{\text{min}} < t < t_{\text{max}}$.

To implement the lookups in OptiX, we have to set up a programmable OptiX pipeline.
The pipeline consists of multiple user-provided functions, called \textit{programs}, as well as some additional configuration options, and it can be launched similar to a CUDA kernel from the host CPU. 
When we start the pipeline, it spawns a CUDA thread for each lookup, where each thread calls the \textit{ray generation program}.
Therein, we concurrently convert each lookup range into the two ray parameters $o$ and $d$, then pass those to the \smalltt{optixTrace()} API function to initiate the hardware-accelerated tracing procedure.
Precisely, for range lookup~\mbox{$[l^{(i)}$, $u^{(i)}]$}, \smalltt{optixTrace()} receives $o = (l^{(i)} - 0.5, 0, 0)$, $d = (1, 0, 0)$, $t_{\text{min}} = 0$, and $t_{\text{max}} = u^{(i)} - l^{(i)} + 1$, along with a reference to the pre-computed BVH.
To obtain intersection information, we also have to define the so-called \textit{any-hit program}, which is called when the tracing procedure finds a ray-triangle intersection.
The any-hit program receives the offset of the triangle within the vertex buffer, which corresponds to the rowID in the original table, for further processing.

\vspace*{-0.1cm}
\section{Design Choices}
\label{sec:choices}
\vspace*{-0.1cm}

After discussing the core principle, let us discuss five fundamental design choices we face.
We carefully evaluate all options to identify the strengths and weaknesses of each choice.
Before that, let us introduce our evaluation setup.

\begin{table*}[t]
\centering
\begin{tabular}{|C{1.5cm}||C{2.5cm}|L{6cm}|C{2cm}|c|c|c|}
\toprule
Mode & Distinct Keys & Conversion Formula & Gap Creation & Triangles & Spheres & AABBs \\
\midrule
\textbf{Naive} & $2^{23}$ & $k \mapsto (\text{\smalltt{float}}(k), 0.0, 0.0)$ & $\pm 0.5$ & \yes & \yes & \yes  \\
\textbf{Extended} & $2^{29}$ & $k \mapsto (\text{\smalltt{bit\_cast<float>}}(2k + C), 0.0, 0.0)$ & \smalltt{nextafter()} & \yes & \no & \yes  \\
\textbf{3D} & $2^{64}$ & $k \mapsto (\text{\smalltt{float}}(k_{22:0}), \text{\smalltt{float}}(k_{45:23}), \text{\smalltt{float}}(k_{63:46}))$ & $\pm 0.5$ & \yes & \yes & \yes  \\
\bottomrule
\end{tabular}
\caption{Overview of our proposed order-preserving methods for converting integers to floating-point numbers.}
\vspace*{-0.3cm}
\label{tab:conversion}
\end{table*}

\vspace*{-0.2cm}
\subsection{Experimental Setup and Methodology}
\label{ssec:choices_setup}
\vspace*{-0.1cm}

As we purely target GPU-resident data management, which becomes increasingly attractive due to an increase in available GPU memory, we assume there exists an array containing our key set in GPU memory.
From this array, we construct the actual index, where each key's rowID is determined by its position in the array. 
Looking up a key in the index returns a set of rowIDs, which we subsequently use to retrieve values from a second GPU-resident array of the same size.
This simulates the typical usage of a secondary index.
As a final result, we compute the sum of all retrieved values. 
In our evaluation, we perform both \textit{point lookups}, where we look up an exact key~$k$ in the index, as well as \textit{range lookups}, where we look up all keys within a range~$[l, u]$.
As mentioned, we always perform batch lookups to utilize the parallel nature of the hardware.
In this case, all results for the batch of lookups are stored in a corresponding result array.
Note that if a lookup does not return any rowIDs, a reserved \textit{miss value} is written into the result array instead.

In our initial set of experiments, we fill the key array with $2^{26}$ consecutive unsigned 32-bit integers, starting at zero, where the keys are shuffled arbitrarily.
We use a dense key set here to ensure a predictable number of hits -- later on in Section~\ref{sec:experiments}, we will evaluate sparse key sets, duplicate keys, and varying hit rates as well.
To generate the point lookups, we uniformly and randomly choose keys from the key array.
For range lookups, we also uniformly pick a lower bound~$l$ from the key array and increase it by the desired number of hits to generate the upper bound~$u$.
In total, we generate $2^{27}$ lookups for every experiment and fire them in a single batch unless specified otherwise.
As we will see in the evaluation, only batch processing workloads, which, for instance, arise naturally in index-based joins, are able to fully saturate the GPU.
We always report the average of five runs (after an initial ``warmup run'' to check for correctness and to ensure that the GPU does not wait for the release of resources anymore).
Regarding hardware, our system contains an NVIDIA RTX~4090 GPU with $24$~GB of VRAM and $128$~raytracing cores.
This GPU implements the most recent Ada Lovelace architecture and is the fastest consumer RTX GPU currently available. 
In Section~\ref{ssec:hardware}, we compare the performance with three other GPUs of two older RTX architectures.

Note that whenever experiments require a deeper investigation, we use the following two GPU profiling tools:
NVIDIA's \textit{Nsight Systems}~\cite{lit:nsight_systems} tool can visualize the order and run time of GPU activities, such as kernel launches and memory allocations.
Another tool, \textit{Nsight Compute}~\cite{lit:nsight_compute}, provides detailed hardware metrics for individual kernels and for the user-programmable parts of the raytracing pipeline.
Unfortunately, Nsight Compute does not provide a cost breakdown for the fixed-function parts of the raytracing pipeline.
However, we ran experiments to ensure all memory counters, which we frequently rely on, cover the pipeline end-to-end.

\vspace*{-0.3cm}
\subsection{How Can We Express Keys?}
\label{ssec:key_range}
\vspace*{-0.1cm}

The first configuration dimension centers around how we can express our keys in a 3D scene.
This question is more challenging than it seems since the straightforward implementation we described so far omitted an inconvenient detail:
OptiX only supports single-precision floating-point numbers (\smalltt{float32}) to represent 3D coordinates.
As a consequence, simply casting a 32-bit integer key (or even a 64-bit key) to a floating-point coordinate will result in a loss of precision, and therefore, wrong results. 
In the following, we propose three different options to still express keys.
With each presented option, we extend the supported key range up to 64-bit. 

\textbf{Naive Mode}. 
We start with the Naive Mode, where the high-level idea is to naively express an integer key as a \smalltt{float32} vertex coordinate.
To understand the problem with this approach, let us look at how the integer~$22$ can be represented as a \smalltt{float32}.
The binary representation of $22$ is $(10110)_2$ or $(10110.0)_2$ with a binary point.
To store this number as a \smalltt{float32}, we first convert it into its normalized form, which means the binary point is shifted next to the most significant bit.
In our example, this results in a shift of four~positions, which we can express as
$(10110.0)_2 = 2^{\textcolor{FloatExponent}{4}} \times \textcolor{FloatSignificand}{(1.011)_2}\text{.}$  
Here, \textcolor{FloatExponent}{$4$} is called the \textcolor{FloatExponent}{exponent~$e$}, whereas \textcolor{FloatSignificand}{$(1.011)_2$} is called the \textcolor{FloatSignificand}{significand~$m$}.
In a \smalltt{float32}, the (signed) exponent is represented using $8$ bits, whereas the significand can have at most $24$ bits. 
Unfortunately, with a significand of $24$ bits, the contiguous range of non-negative integers that can be stored in a \smalltt{float32} is at most $2^{24}$.
However, the situation is even worse: 
OptiX requires us to leave a gap between the start/end of the ray and the adjacent triangles for the hit to be registered (see Figure~\ref{fig:core_principle}).
Thus, we also need to make sure that for each key $k$, $k \pm 0.5$ can be represented as a \smalltt{float32}.
Therefore, we have to conservatively restrict the key range to $2^{23}$.
Only then, querying the very last key $2^{23}-1$ will work, as we can express $t_{max} = 2^{23}-1+0.5$ without a loss of precision.
This would not be the case for $2^{24}-1$, where $t_{max} = 2^{24}-1+0.5$ cannot be represented.


\begin{figure}[!ht]
\centering
\begin{subfigure}[b]{.9\linewidth}
\includegraphics[width=1.0\linewidth]{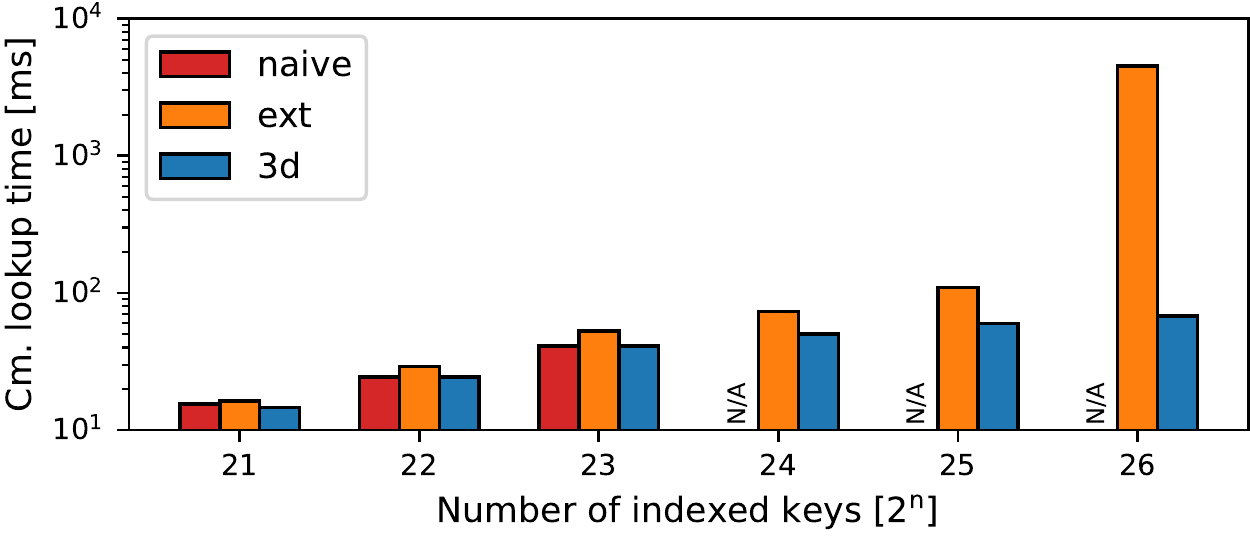}
\caption{Standard conversion.}
\label{fig:keyconv:standard}
\end{subfigure}
\begin{subfigure}[b]{.9\linewidth}
\includegraphics[width=1.0\linewidth]{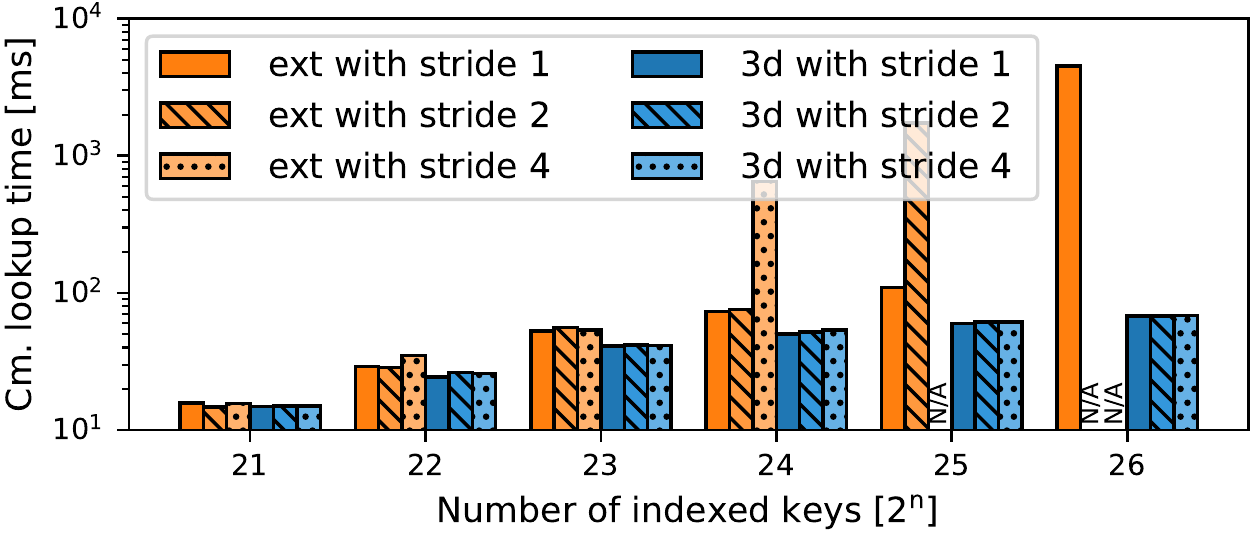}
\caption{Introducing stride.}
\label{fig:keyconv:stride}
\end{subfigure}
\caption{Effects of key representations on lookup time. Notice the irregular behavior of Extended Mode (``ext'').}
\vspace*{-0.2cm}
\label{fig:keyconv}
\end{figure}

\textbf{Extended Mode}. 
So far, we have converted each integer key~$k$ to its corresponding \smalltt{float32} representation.
This limited our supported key range to $2^{23}$.
However, the range of floating point numbers that can be represented by a \smalltt{float32} is significantly larger than $2^{23}$.
To exploit this larger range, we need an order-preserving mapping from an integer key~$k$ to a corresponding floating point number~$f$. 
We propose the following mapping:
Each integer key~$k$ is mapped to the $2k$-th representable floating-point number using \smalltt{bit\_cast<float>(2k)}.
Mapping to every second \smalltt{float32} ensures that there is always a "gap value" between adjacent keys.
The gap values next to a key $k$ can be identified by passing $k$ to the \smalltt{nextafter()} function from the C standard library (instead of computing $k \pm 0.5$).
Finally, note that we offset $2k$ by a constant~$C$ prior to casting, as not offsetting yields wrong results due to implementation details related to \smalltt{float32} processing.
We found $C = \smalltt{bit\_cast<uint32\_t>(0.5f)}$ to produce correct results for all keys up to $2^{29}$.

\textbf{3D Mode}.
While we can already express $2^{29}$ distinct keys in Extended Mode, a general-purpose index structure should be able to operate with $64$-bit keys. To achieve this, and since all vertices in OptiX are three-dimensional anyway, we now decompose the key bits into three smaller integers. We covert these integers to \smalltt{float32} individually, and use them as three-dimensional coordinates to center each triangle around.
In our case, for a 64-bit key~$k$, we use the \textbf{$23$}~least significant bits as the $x$ coordinate (written as $x = k_{22:0}$), the next $23$~bits constitute the $y$ coordinate, and the remaining $18$~bits form the $z$ coordinate.
In Section~\ref{ssec:decomposition}, we evaluate other decompositions as well.
To support 32-bit keys using this method, we extend each key to $64$~bits by padding it with zeros.
This mode is identical to Naive Mode for all keys smaller than~$2^{23}$.

\begin{figure}[!ht]
\centering

\begin{subtable}[b]{0.46\linewidth}
\centering
{
\small
\begin{tabular}{|c|c|c|c|}
\toprule
\# & $k$ & $x_k = k_{1:0}$ & $y_k = k_{63:2}$ \\
\midrule
\textcolor{Qualifies}{0} & \textcolor{Qualifies}{19} & 3 & 4 \\
\textcolor{Qualifies}{1} & \textcolor{Qualifies}{17} & 1 & 4 \\
2 & 23 & 3 & 5 \\
3 & 14 & 2 & 3 \\
4 & 12 & 0 & 3 \\
\textcolor{Qualifies}{5} & \textcolor{Qualifies}{15} & 3 & 3 \\
\textcolor{Qualifies}{6} & \textcolor{Qualifies}{20} & 0 & 5 \\
\bottomrule
\end{tabular}
}
\caption{Keys.}
\label{fig:3d:keys}
\end{subtable}
\begin{subfigure}[b]{0.53\linewidth}
\includegraphics[page=1, width=\linewidth, trim={1mm 47mm 252mm 84mm}, clip]{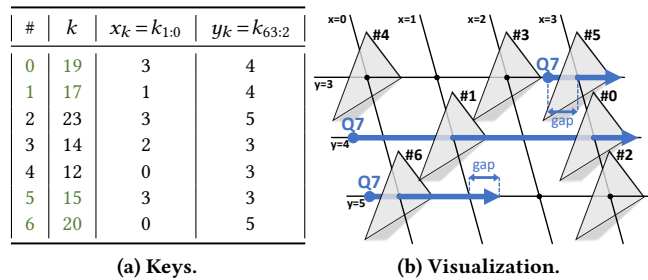}
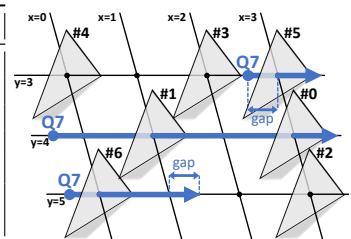
\caption{Visualization.}
\label{fig:3d:vis}
\end{subfigure}
\caption{Answering the range lookup [15,21] in 3D Mode (simplified to two dimensions).}
\label{fig:3d}
\vspace*{-0.2cm}
\end{figure}

Note that this approach requires slight modifications to point lookups and range lookups. 
For point lookups, the origin~$o$ now requires a three-dimensional offset. 
For range lookups, a single ray might now be insufficient, since the triangles do not form a single "line" anymore, but are scattered across the 3D scene.
Instead, we have to cast distinct rays for each integer between $l_{63:23}$ and $u_{63:23}$.
If a range lookup spans at most $2^{23}$ integers, it can be answered by casting only one or two rays.
Figure~\ref{fig:3d} demonstrates how to answer the range lookup $[l,u] = [15,21]$ (\textbf{Q4}) in 3D Mode.
To visualize the principle in the example, we assume to have only two dimensions, where the $x$ coordinate is determined by the two least significant bits of the corresponding key, and the $y$ coordinate by all remaining bits.
Table~\ref{fig:3d:keys} shows the indexed keys along with their $x$ and $y$ coordinates in decimal representation.
We first split the upper and lower bounds into their coordinates, i.e., $l = 15$ into $x_l=l_{1:0}=3$ and $y_l=l_{63:2}=3$, and $u = 21$ into $x_u=u_{1:0}=1$ and $y_u=u_{63:2}=5$.
As the range of interest along the $y$-axis ranges from 3 to 5, we have to fire three corresponding rays in parallel to the $x$-axis.
The first ray we fire starts at $x_l - 0.5=2.5$, the last ray ends at $x_u + 0.5 = 1.5$, whereas all intermediate rays (for $y=4$ in this example) are unbounded, i.e., they hit all triangles along that line.

Let us now see how the three key conversion methods perform in comparison.
Figure~\ref{fig:keyconv:standard} shows the cumulative lookup times when varying the build size from $2^{21}$ to $2^{26}$.
As we can see, the lookup times are very similar between all three conversion methods, apart from one oddity:
With Extended Mode, lookup takes an extraordinary amount of time as soon as the build size exceeds $2^{25}$.
One could suspect that the \textit{magnitude of keys} might be responsible, since the numbers produced in Extended Mode become very large at some point, up to around $2^{62}$/$10^{18}$.
We tested this suspicion by multiplying each key by $\pm 2^{20}$ after conversion, but for all conversion methods, the lookup time was identical.
Instead, the \textit{value range of keys} appears to be responsible for this behavior (i.e., the ratio $q$ between the largest and the smallest inserted key), and we introduced \textit{key stride} to verify this:
Instead of inserting keys $1$, $2$, $3$, etc., we insert $1s$, $2s$, $3s$, etc., with stride parameter $s=1$, $s=2$, and $s=4$ in Figure~\ref{fig:keyconv:stride}.
The lookup times increase drastically as soon as $q$ hits $2^{26}$, and for any larger $q$, the experiment timed out after five minutes.
This is confirmed by profiling, which reveals that in Extended Mode, the raytracing pipeline loads $76\times$ more data from main memory, and $93\times$ more data from the internal L2 cache when compared to 3D Mode, despite the build/lookup setup being identical.
This indicates that the BVH has to traverse more triangles internally, implying that less triangles could be excluded outright, and cache bandwidth becomes the bottleneck.

\textbf{Selected Configuration}. In conclusion, only 3D Mode can represent 64-bit keys, and it also exhibits stable scaling behavior. Therefore, we will use this mode as the default key conversion method for all subsequent experiments.

\textbf{Handling other data types}.
Before continuing, we want to emphasize that \textbf{RX} does not only support unsigned 64-bit integers, but can handle other data types as well.
All native C data types can be mapped to a \smalltt{uint64} while preserving their relative order (this technique is traditionally used in radix sorting), and can therefore be indexed by \textbf{RX}.
In particular, \smalltt{float32} and \smalltt{float64} values should always be converted in this way, and never be indexed directly, since the ratio $q$ between the smallest and largest value might be very large, leading to immense slowdowns.
Composite data types (structs, arrays, strings) can still benefit from \textbf{RX} if their natural ordering is lexicographic.
For these data types, we can convert the first few components (e.g., the first eight characters in a string) into unsigned integers individually, then densely pack them into a single 64-bit integer that can be passed to \textbf{RX}.
This results in hardware-accelerated point and range lookups for the first 64~bits of the data type.
The remaining components have to be compared and filtered in software.

\vspace*{-0.2cm}
\subsection{How Should We Cast Rays for Lookups?}
\label{ssec:rays}
\vspace*{-0.1cm}

Next, let us discuss and evaluate the options we have in casting rays for range and point lookups.
Let us first look at answering a range lookup $[l, u]$, which also covers point lookups by setting~$l=u$.

\begin{figure}[!ht]
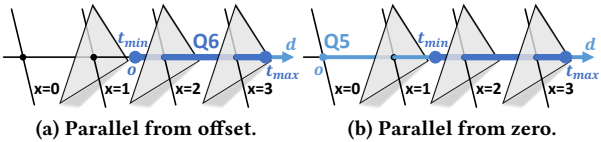

\centering
\begin{subfigure}[b]{0.48\linewidth}
\includegraphics[page=1, width=\linewidth, trim={248mm 117mm 23mm 50mm}, clip]{figures/rays}
\caption{Parallel from offset.}
\label{fig:ray-origin:offset}
\end{subfigure}%
\begin{subfigure}[b]{0.48\linewidth}
\includegraphics[page=1, width=\linewidth, trim={164mm 117mm 107mm 50mm}, clip]{figures/rays}
\caption{Parallel from zero.}
\label{fig:ray-origin:zero}
\end{subfigure}%
\caption{Two ways of expressing the range lookup [2, 3].}
\vspace*{-0.2cm}
\label{fig:ray-origin}
\end{figure}

Figure~\ref{fig:ray-origin} shows the two options we have to look up the range~$[2, 3]$:  
In \textbf{Parallel from offset} (Figure~\ref{fig:ray-origin:offset}), which we used so far, the ray~(\textbf{Q5}) originates at~$l-0.5$ and ends at $u+0.5$. Thus, the origin of the ray is offset from zero.
In \textbf{Parallel from zero} (Figure~\ref{fig:ray-origin:zero}), the ray~(\textbf{Q6}) always originates from $0$ and ends after $u+0.5$. To avoid false positives, the ray parameter $t_{\text{min}}$ is set to $l-0.5$. 
For point lookups, we have a third option available:
Using \textbf{Perpendicular} (Figure~\ref{fig:core_principle}), the ray always targets only one specific triangle and is fired from a perpendicular angle to it. 
Table~\ref{tab:rays} summarizes the three options with their respective ray parameters.
Note that in 3D Mode, we additionally shift the origin to the correct $y$ and $z$ coordinates (see Section~\ref{ssec:key_range}).

\begin{table}[!ht]
\centering
{
\small
\begin{tabular}{|L{2.5cm}|c|c|c|c|}
\toprule
Method & $o$ & $d$ & $t_{\text{min}}$ & $t_{\text{max}}$ \\
\midrule
\textbf{Para. offset} (\textbf{Q5}) & $(l - 0.5, 0, 0)$ & $(1, 0, 0)$ & $0$ & $u - l + 1$ \\
\textbf{Para. zero} (\textbf{Q6}) & $(0,0,0)$ & $(1,0,0)$ & $l - 0.5$ & $u + 0.5$ \\
\midrule
\textbf{Perpend.} (\textbf{Q3}) & $(l, 0, - 0.5)$ & $(0, 0, 1)$ & $0$ & $1$ \\
\bottomrule
\end{tabular}
}
\caption{Ray configuration parameters.}
\vspace*{-0.3cm}
\label{tab:rays}
\end{table}

\begin{figure}[!ht]
\centering
\includegraphics[width=.9\linewidth]{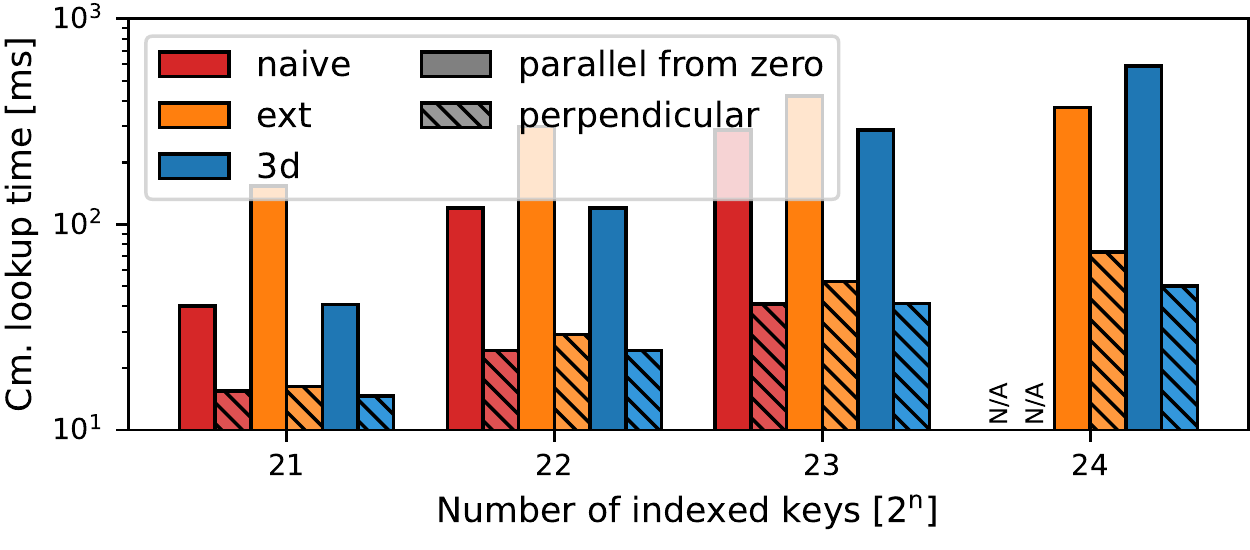}
\caption{Lookup time for parallel and perpendicular rays.}
\vspace*{-0.3cm}
\label{fig:parallel-vs-perpendicular-time}
\end{figure}

In Figure~\ref{fig:parallel-vs-perpendicular-time}, we first evaluate whether point lookups should be expressed as parallel rays or as perpendicular rays.
Note that Extended Mode does not support offsetting the ray origin due to \smalltt{float32} precision limits.
Therefore, we only test rays starting from zero in this experiment.
We can clearly see that perpendicular rays consistently yield better lookup times than parallel rays.
This is because the parallel ray, by definition, intersects with the majority of the bounding boxes and has to rely on $t_{min} < t < t_{max}$ to exclude potential hits, while the perpendicular ray misses most bounding boxes by default.

To identify which choice works better for range lookups, in Table~\ref{tab:ray-origin-time}, we compare the answering time for offset rays, and rays starting at zero. 
Note that we evaluate only 3D Mode here, since Extended Mode does not support rays with offset, and Naive Mode works almost identically to 3D Mode.
We can see that offsetting the origin pays off in all cases over keeping the origin at zero. 

\textbf{Selected Configuration.} \textbf{RX} uses perpendicular rays for point lookups and rays with an offset origin for range lookups. 

\begin{table}[!ht]
\centering
{
\small
\begin{tabular}{|l|C{0.7cm}|C{0.7cm}|C{0.7cm}|C{0.7cm}|C{0.7cm}|}
\toprule
Number of hits & $1$ & $4$ & $16$ & $64$ & $256$ \\
\midrule
Parallel from offset [ms] & 61 & 197 & 580 & 2086 & 8025 \\
Parallel from  zero [ms] & 61 & 1209 & 1652 & 3382 & 10196 \\
\bottomrule
\end{tabular}
}
\caption{Lookup time for two choices of ray origin for range lookups in 3D~Mode.}
\vspace*{-0.3cm}
\label{tab:ray-origin-time}
\end{table}

\begin{figure*}[!ht]
\centering
\begin{subfigure}[b]{0.33\textwidth}
\includegraphics[width=1.0\linewidth]{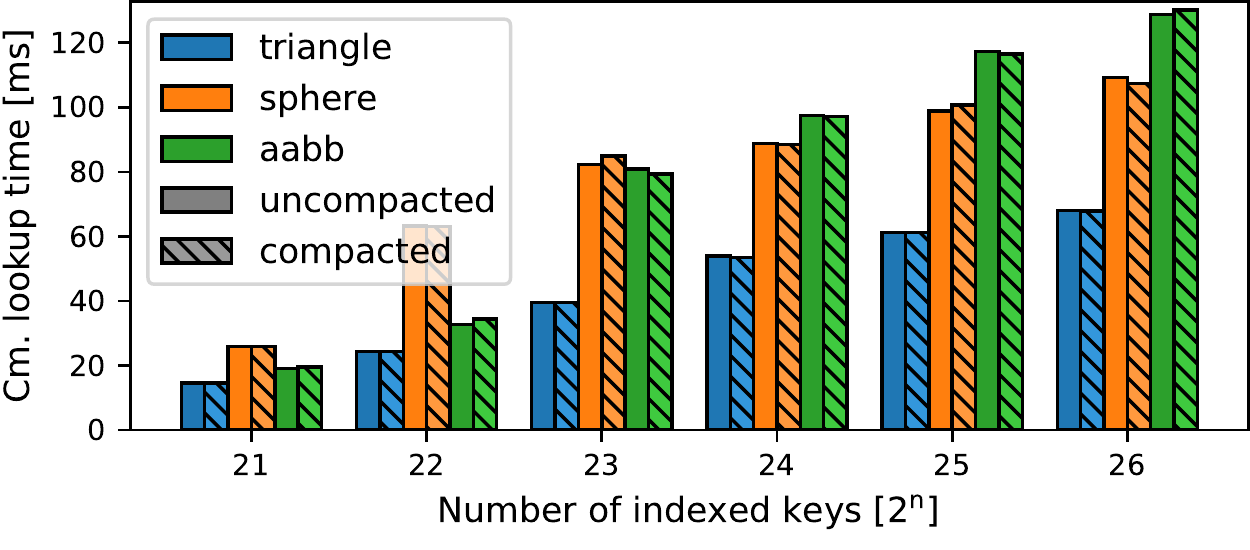}
\caption{Lookup performance.}
\label{fig:primitives:lookup}
\end{subfigure}
\begin{subfigure}[b]{0.33\textwidth}
\includegraphics[width=1.0\linewidth]{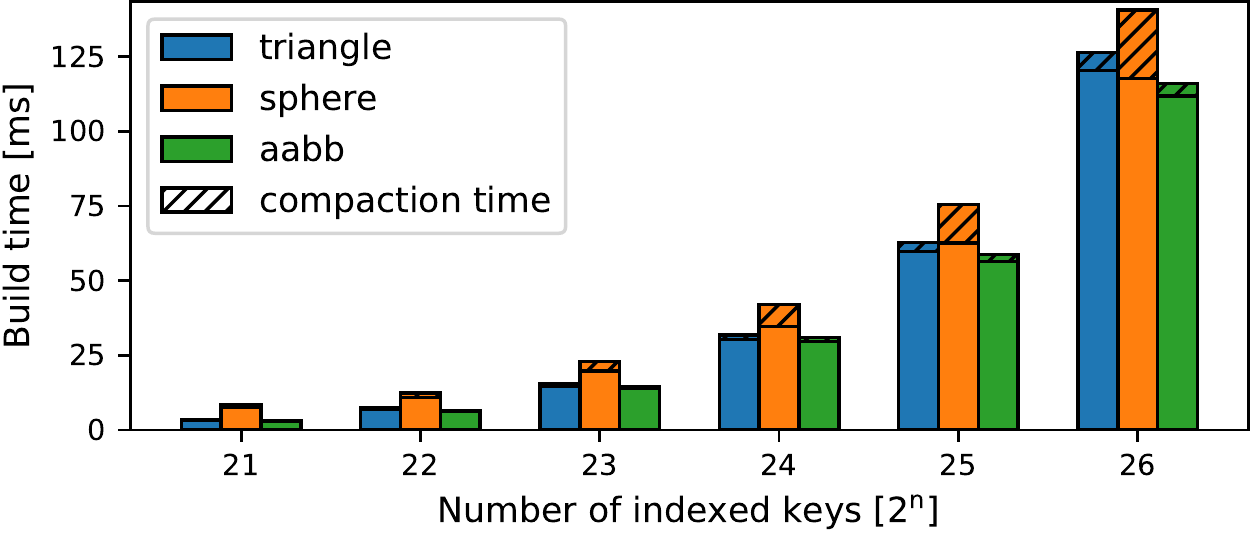}
\caption{Build performance.}
\label{fig:primitives:build}
\end{subfigure}
\begin{subfigure}[b]{0.33\textwidth}
\includegraphics[width=1.0\linewidth]{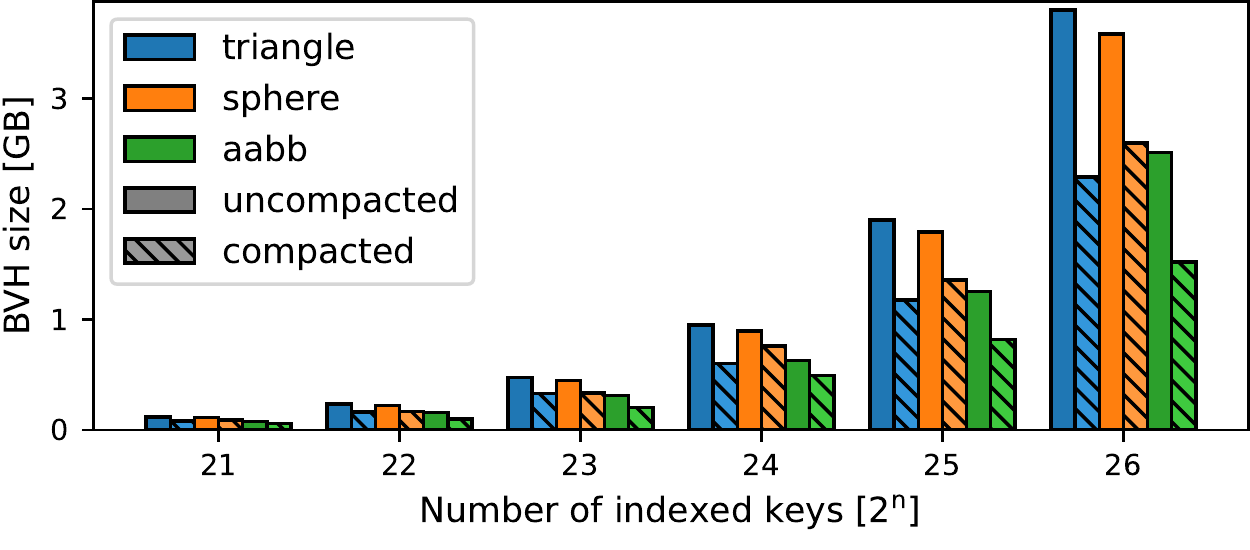}
\caption{Memory footprint.}
\label{fig:primitives:memory}
\end{subfigure}
\caption{Comparison of primitive types.}
\label{fig:primitives}
\end{figure*}

\vspace*{-0.5cm}
\subsection{How Can We Decompose the Key?}
\label{ssec:decomposition}

\begin{figure}[h]
\vspace*{-0.2cm}
\includegraphics[width=.95\linewidth]{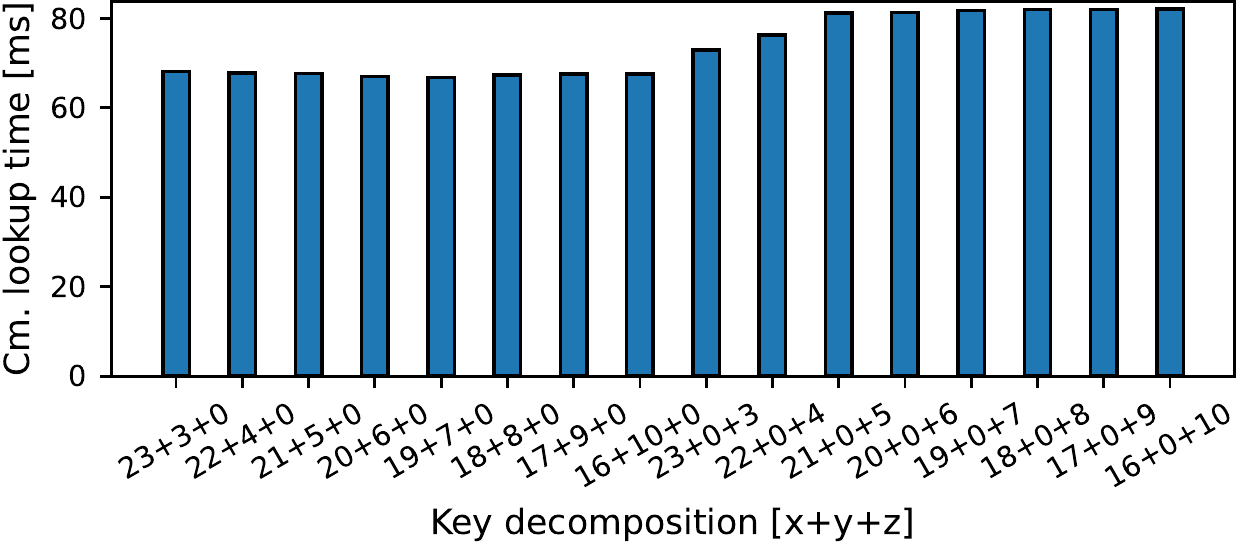}
\caption{Point lookups under varying key decompositions.}
\vspace*{-0.3cm}
\label{fig:decomp_point_lookups}
\end{figure}

In Section~\ref{ssec:key_range}, we have decomposed each key~$k$ in 3D Mode as $x = k_{22:0}$, $y=k_{45:23}$, and $z=k_{63:46}$.
This decomposition allowed us to support 64-bit keys.
Since other decompositions are possible as well, in the following, we test different configurations to see their impact on the performance.
From Figure~\ref{fig:decomp_point_lookups}, we can see that the choice of decomposition affects lookup performance.
When we allocate bits to the $x$ and $z$ components (bars on the right side), the lookup time increases.
Remember that we always fire rays along the $z$-axis to answer point lookups.
Assigning more bits to the $z$ component means that triangles will increasingly stack along the $z$-axis, which effectively turns the perpendicular ray into a parallel ray (see Section~\ref{ssec:rays}).
In contrast, when all triangles satisfy $z=0$ (bars of the left side), there is only one layer of triangles from the perspective of the ray, and lookups are fast.
Note that we also tested a variant where the keys are not densely packed, but uniformly picked from the entire 64-bit space.
In this situation, as expected, the lookup time was not affected by the decomposition at all, since there is no dimensional clustering present in the key set.

\begin{figure}[h]
\includegraphics[width=.95\linewidth]{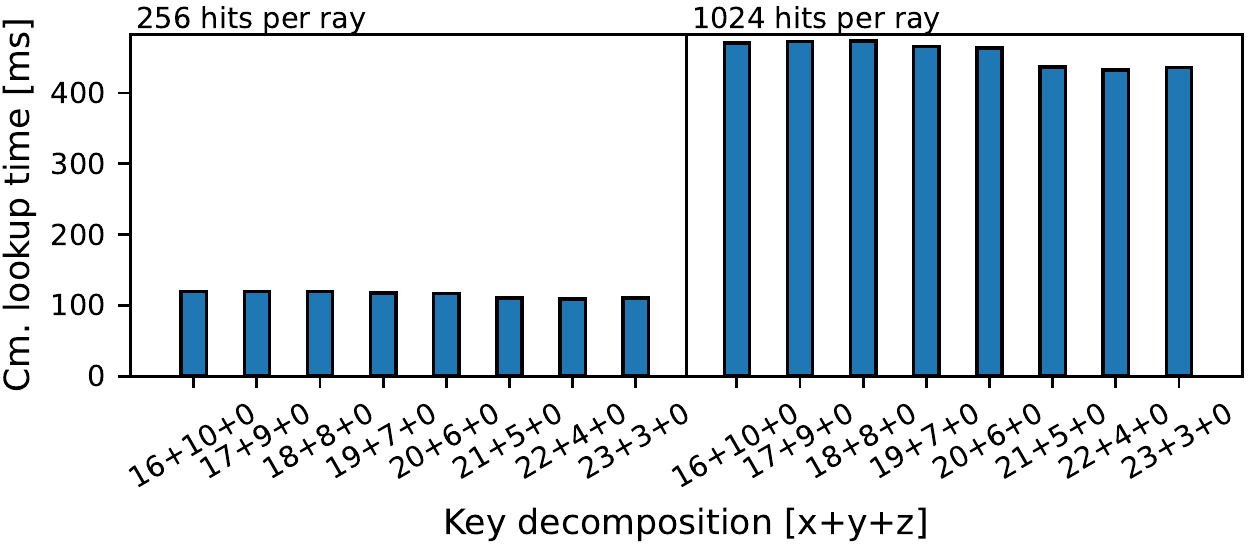}
\caption{Range lookups under varying key decompositions.}
\vspace*{-0.5cm}
\label{fig:decomp_range_lookups}
\end{figure}

In Figure~\ref{fig:decomp_range_lookups}, we additionally evaluate different decompositions for range lookups.
We can see that the more the bits are assigned to the $x$-dimension, the better the lookup time.
Maximizing the number of bits for the $x$-component also reduces the risk of having to cast multiple rays for wide ranges (see Figure~\ref{fig:3d:vis}), and if multiple rays must be cast, their number is minimized.

\textbf{Selected Configuration.} \textbf{RX} uses the decomposition $x = k_{22:0}$, $y=k_{45:23}$, and $z=k_{63:46}$ throughout the rest of the paper, which yields good results for both point and range lookups.  

\vspace*{-0.2cm}
\subsection{Which Primitive Type Is Ideal?}
\label{ssec:primitives}
\vspace*{-0.1cm}

So far, we have discussed how to express our keys using triangles.
As each triangle is stored as nine \smalltt{float32} (three vertex coordinates for each of the three vertices), let us see whether other primitive types offer a better memory footprint and/or performance.
Apart from triangles, OptiX supports \textit{spheres} and \textit{axis-aligned bounding boxes (AABBs)}.

\textbf{Spheres}. In contrast to triangles, which are planar, spheres are curved surfaces defined by their center point and their radius.
A ray-sphere intersection can only occur when the ray enters or exits the volume of the sphere at some point.
To ensure that a ray can always start outside of a sphere, we must avoid packing the spheres too densely.
Just like with triangles, rays have to start and end in the gaps between adjacent spheres.
To leave a sufficiently large gap, we choose $r = 0.25$ as the radius for each sphere.
Since the radius is uniform for all spheres, OptiX allows us to specify the radius for all spheres at once.
Consequently, each sphere only requires three \smalltt{float32} to store the center, making this representation comparatively space-efficient.

\textbf{AABBs}. Like spheres, AABBs delimit a volume, and only register an intersection when a ray strikes one of the six faces.
AABBs are intended to enclose user-defined primitives, e.g., implicitly defined surfaces, allowing them to be part of a larger 3D scene without requiring them to be supported by OptiX natively.
Consequently, the user is expected to provide their own \textit{intersection program} to figure out if a ray actually hits the object enclosed by the bounding box.
In our case, it is sufficient to move the contents of the any-hit program into the intersection program, and not report a hit in the end.
Internally, each AABB is represented by two corner points on opposite sides, requiring six \smalltt{float32} in total and making AABBs more space-efficient than triangles.

Let us now compare the cumulative lookup time, the build time, and the memory footprint for all three primitive types in Figure~\ref{fig:primitives}.
We show the results for both the (default) uncompacted variant, as well as when an additional compaction step via \smalltt{optixAccelCompact()} is performed afterwards.
Regarding lookup performance, triangles clearly perform best with a significant margin.
We attribute this to the fact that the ray-triangle intersection test is implemented in hardware~\cite{lit:optix-turing}, utilizing the raytracing cores, whereas spheres and AABBs both call a software-based intersection program.
Both the uncompacted and the compacted variant perform almost identically for all primitive types. 
In terms of BVH build time, AABBs perform best, closely followed by trianges, while sphere BVHs take longer to create.
Compacting the structure after building it is cheap for all three methods, where especially for triangles, the overhead is negligible.
Unfortunately, the memory footprint of (uncompacted) triangles is the highest in comparison. At the same time, we can see that the memory footprint decreases by up to 50\% under compaction. 
Surprisingly, a sphere BVH takes the most time to compact, and the final memory footprint is the largest of the three primitives. 

\textbf{Selected Configuration}. When optimizing for lookup performance, triangles should be preferred in order to fully utilize hardware acceleration. When optimizing for memory footprint, AABBs should also be considered. As we prioritize lookup performance, \textbf{RX} uses triangles. Also, we compact the BVH in all cases. 

\begin{figure*}[!ht]
\centering
\begin{subfigure}[b]{0.33\textwidth}
\includegraphics[width=\linewidth]{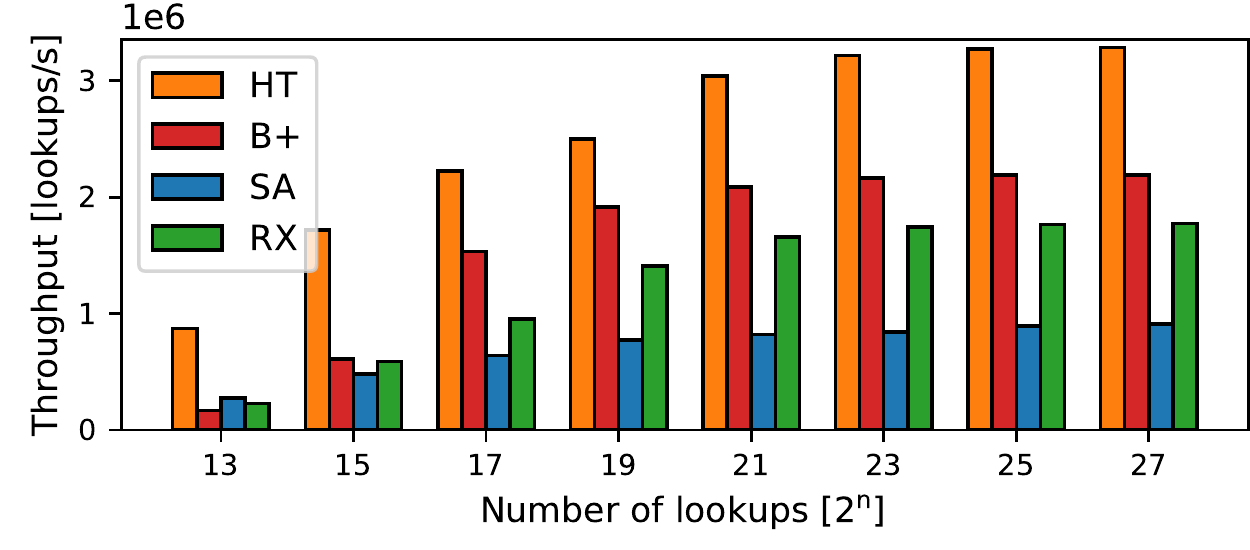}
\caption{Throughput (varying number of lookups).}
\label{fig:probe_size}
\end{subfigure}
\begin{subfigure}[b]{0.33\textwidth}
\includegraphics[width=\linewidth]{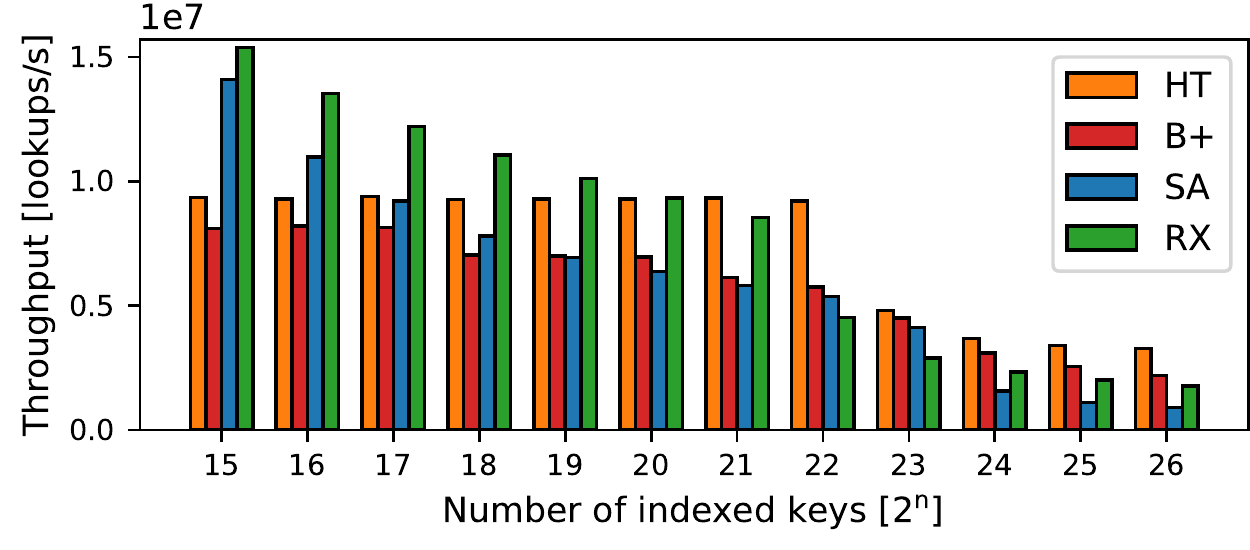}
\caption{Throughput (varying build set size).}
\label{fig:varying_build_size:throughput}
\end{subfigure}
\begin{subfigure}[b]{0.33\textwidth}
\includegraphics[width=\linewidth]{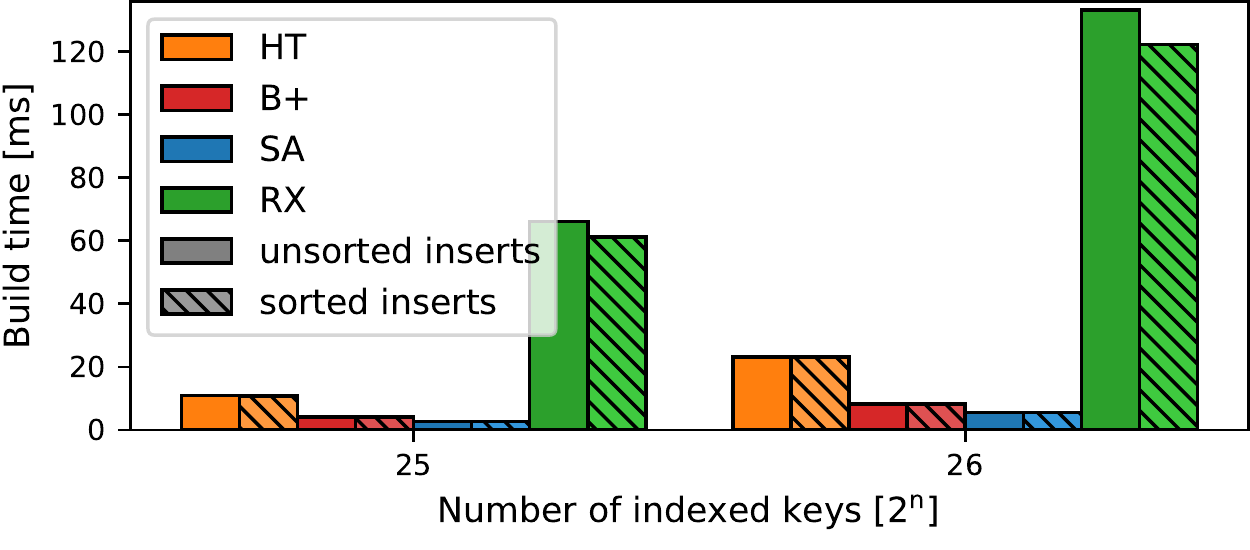}
\caption{Impact on the build time.}
\label{fig:varying_build_size:build_time}
\end{subfigure}
\caption{The scaling behavior of all indexing methods.}
\vspace*{-0.3cm}
\label{fig:varying_build_size}
\end{figure*}

\vspace*{-0.1cm}
\subsection{How Do We Perform Updates?}
\label{ssec:updates}
\vspace*{-0.1cm}

Next, let us discuss how we perform updates on an already existing index. 
OptiX natively supports in-place updates to bounding volume hierarchies via \smalltt{optixAccelBuild()}, albeit with a few restrictions~\cite{lit:optix-guide-updates}:
A special flag has to be set during construction, which disables the effects of compaction.
Further, updates still require additional temporary memory, and updates cannot add new primitives or remove existing primitives. 

To evaluate OptiX' update behavior, we build the BVH as described earlier and set the aforementioned update flag.
We then test two different update workloads:
(a)~We swap pairs of adjacent positions in the buffer. Since keys are not sorted in the buffer, this simulates updates that significantly change keys. 
(b)~We swap pairs of (rank-)adjacent keys. This simulates updates that change keys by $\pm 1$, since our key set is dense.
In both cases, the set of keys itself does not change, only the keys' position within the buffer changes.
As such, we would expect the lookup time to remain identical.
After applying updates to the key buffer, we again convert all keys into triangles and update the BVH via \smalltt{optixAccelBuild()}.
Note that the triangle buffer does not only contain the updated primitives, but also the untouched ones.
After the updates have been applied, we perform the usual lookup phase.

\begin{table}[!ht]
\centering
{
\small
\begin{tabular}{|L{1.6cm}|l||c|c|c|c|c|}
\toprule
Experiment & Phase & $2^4$ & $2^8$ & $2^{12}$ & $2^{24}$ & rebuild \\
\midrule
\multirow{3}{*}{\shortstack[l]{Swap adj.\\ positions}}
& Updates & 38.7 & 38.8 & 38.7 & 39.6 & 126.5 \\
& Lookups & 68.1 & 129.1 & 5361.3 & - & 68.1 \\
& Total & 106.8 & 167.9 & 5400.0 & - & 194.6 \\
\midrule
\multirow{3}{*}{\shortstack[l]{Swap adj.\\ keys}}
& Updates & 39.5 & 39.5 & 39.5 & 39.5 & 126.5 \\
& Lookups & 68.1 & 68.2 & 68.2 & 68.2 & 68.1 \\
& Total & 107.6 & 107.7 & 107.7 & 107.7 & 194.6 \\
\bottomrule
\end{tabular}
}
\caption{Update and lookup time in milliseconds for $2^{26}$ keys when swapping adjacent buffer positions and adjacent keys.}
\vspace*{-0.3cm}
\label{fig:update-time}
\end{table}

From Table~\ref{fig:update-time}, we can make a set of interesting observations:
(1)~The time required to update the structure is independent from the number of applied updates.
This is the case since the entire buffer must be passed to the update routine -- not only the updated entries.
(2)~Updating is still more than three times faster than rebuilding the BVH from scratch.
This suggests that the BVH is not completely rebuilt, but the existing bounding volumes are merely adjusted.
(3)~This adjustment can drastically impact the lookup time.
When swapping more than $2^{8}$ keys in adjacent buffer positions, the adjusted bounding volumes are much larger than before, increasing the amount of required intersection tests.
In such a situation, a full rebuild should be preferred.
This behavior of adjusting the BVH is confirmed by the stable lookup times when swapping rank-adjacent keys instead of adjacent positions.
A profiler can provide more insight via memory statistics:
Although the amount of main memory accesses does not differ significantly between the two update variants, swapping adjacent positions leads to an immense increase in L1/L2 cache reads, while the increase is barely noticeable when swapping adjacent values.
Again, this indicates that the BVH cannot exclude as many potential hits as before.
The OptiX documentation confirms this behavior~\cite{lit:optix-guide-updates}, claiming that the quality of the BVH will degrade substantially when too many triangles are relocated.

\textbf{Selected Configuration}. We conclude that updates in \textbf{RX} should be realized via a full rebuild in favor of lookup performance.

\vspace*{-0.1cm}
\section{Experimental Evaluation}
\label{sec:experiments}
\vspace*{-0.1cm}

After identifying a reasonable and well-performing configuration, we will now vary the experimental setup along nine dimensions and compare \textbf{RX} against three traditional GPU-resident index structures as baselines.
The experimental setup in this section is the same as in Section~\ref{ssec:choices_setup}, however, instead of restricting the key set to consecutive integers, we now permit the full 32-bit integer range as keys (the B+-Tree baseline does not support 64-bit keys).
Any changes to this setup will be stated explicitly.

\vspace*{-0.2cm}
\subsection{Traditional GPU Indexes as Baselines}
\label{ssec:baselines}
\vspace*{-0.1cm}

We compare against the following GPU-resident index structures:

\textbf{HT}.
Our first baseline is WarpCore, a state-of-the-art GPU hashtable~\cite{lit:hash-warpcore, lit:hash-warpcore-code}.
WarpCore implements \textit{cooperative probing}, where each key is assigned to a group of threads during inserts or lookups,
and each group accesses neighboring slots in the hashtable.
This accelerates the task of identifying an empty slot for insertion, or discovering the key to be looked up,
while still using the GPU's cache and load-store units to the maximum extent possible.
The authors show that WarpCore outperforms other recent GPU hashtables, such as SlabHash~\cite{lit:hash-slabhash} and cuDPP~\cite{lit:hash-cudpp1, lit:hash-cudpp2},
especially when the majority of slots is occupied.
Just like the authors, we use a target load factor of 0.8 and fix the group size for cooperative probing to 8.
Since there is no bulk-loading for hashtables, we insert each key separately during the build phase.

\textbf{B+}.
Our second baseline is the state-of-the-art GPU B+-Tree by Awad et al.~\cite{lit:btree1} in its recently updated version~\cite{lit:btree2, lit:btree-code}.
This baseline traverses the tree in groups of 16 threads, so that lookups within a node can be done synchronously using warp intrinsics.
The build phase sorts the keys using CUB's key-value \smalltt{DeviceRadixSort}~\cite{lit:cub}, an out-of-place GPU radix sort that is considered the fastest way to sort integers using a CUDA-enabled GPU, then bulk-loads the tree with the sorted key-value pairs.
Unlike \textbf{HT} and \textbf{RX}, the B+-Tree only supports 32-bit keys.
In comparison to a GPU LSM tree~\cite{lit:lsmtree}, the B+-Tree yields better lookup performance, making it an ideal baseline for the read-only benchmarks in this section.
Note that we had to slightly modify the range lookup code for the B+-Tree in order to support efficient aggregation.

\textbf{SA}.
Our third baseline is a sorted array, which we combine with a naive binary search for lookups.
This mimics the access patterns that would occur during lookups in a balanced binary tree, while requiring less overall memory.
Insertions and deletions cannot take place after the index has been constructed.
Still, this index is trivial to implement, and serves as a great baseline for more sophisticated approaches.
Again, we utilize CUB's radix sort to sort the array.

\vspace*{-0.1cm}
\subsection{Varying the Number of Lookups and the Number Of Indexed Keys}
\label{ssec:scaling}
\vspace*{-0.1cm}

We first compare the scaling behavior, where we focus on lookup throughput and build time, along with memory footprint. 

In Figure~\ref{fig:probe_size}, we first vary the total number of point lookups on the $x$-axis from $2^{13}$ to $2^{27}$ while keeping the number of indexed keys constant at $2^{26}$.
In this experiment, \textbf{HT} clearly outperforms all other indexes.
However, \textbf{RX} remains competitive in comparison to the other order-based index structures \textbf{B+} and \textbf{SA}.
We can also observe that the throughput of all methods starts saturating at around $2^{21}$ lookups per batch.
Below that, the workload is too small to fully utilize the GPU resources.
Let us investigate this behavior for \textbf{RX}:
Each processing element on an NVIDIA GPU (called \textit{SM}) executes threads in groups of 32 (called \textit{warps}).
When running \textbf{RX}, a single SM can dynamically schedule up to 16 warps, which allows the GPU to hide memory latencies, similar to hyper-threading in modern CPUs.
Assigning less than 16 warps to an SM means that the SM is more likely to idle while waiting for a memory dependency.
At the same time, all SMs on the GPU are limited by the peak memory bandwidth, which remains under-utilized when the number of memory accesses is low.
Table~\ref{tab:lookup-util} shows the average number of active warps per SM and the percentage of the peak GPU memory bandwidth utilization:
Increasing the number of lookups rapidly saturates the limit of 16 warps per SM, but also approaches the peak bandwidth at the same time, which eventually leads to constant throughput.
Note that our default of $2^{27}$ lookups will always ensure that the GPU resources are fully utilized during each experiment.

\begin{table}[!ht]
\centering
{
\small
\begin{tabular}{|L{3.25cm}|C{0.6cm}|C{0.6cm}|C{0.6cm}|C{0.6cm}|C{0.6cm}|}
\toprule
Number of lookups & $2^{13}$ & $2^{15}$ & $2^{17}$ & $2^{19}$ & $2^{21}$ \\
\midrule
Active warps per SM & 3.89 & 6.68 & 12.46 & 13.79 & 14.25 \\
Memory BW [\% of peak] & 39.61 & 61.36 & 75.12 & 77.97 & 78.96 \\
\bottomrule
\end{tabular}
}
\caption{Average number of active warps per SM and percentage of the peak GPU memory bandwidth utilization.}
\label{tab:lookup-util}
\vspace*{-0.3cm}
\end{table}

In Figure~\ref{fig:varying_build_size:throughput}, we continue by varying the number of indexed keys from $2^{15}$ to $2^{26}$ while keeping the number of lookups constant at $2^{27}$.
We can see that for smaller key sets of up to $2^{19}$~keys, \textbf{RX} shows the best lookup performance of all methods.
With an increase in the number of keys, the throughput of \textbf{RX} unfortunately falls below \textbf{HT} and \textbf{B+}.
Profiling provides an explanation for this behavior:
When the build set is small, all methods read exactly the same amount of GPU main memory during the lookup phase, indicating that all index structures fit into the GPU's L2 cache.
The profiler also shows that the relative performance of each index structure loosely correlates with the total number of executed instructions:
For $2^{15}$ inserted keys, \textbf{RX} requires very few instructions, since the BVH traversal is done in hardware, while \textbf{B+} requires around $40\times$ as many.
On the other hand, when the size of the build set increases beyond $2^{20}$, the index structures no longer fit into the L2 cache, and the lookup performance is now bounded by GPU memory.
\textbf{RX} and \textbf{B+} load a comparable amount of memory, but both load more than \textbf{HT} and \textbf{SA}.
However, \textbf{SA} falls behind due to latency overhead from unfavorable (random) memory access patterns.

In Figure~\ref{fig:varying_build_size:build_time}, we also inspect the scaling of the build time when doubling the number of keys from $2^{25}$ to $2^{26}$ for both a sorted and an unsorted key set.
\textbf{RX} scales linearly in this regard, however, the BVH creation is significantly more expensive than the build phase for the other indexes.
This, in combination with the fact that updates perform poorly (see Section~\ref{ssec:updates}), indicates that \textbf{RX} should be primarily used as a read-only index structure.

\begin{table}[!ht]
\vspace*{-0.1cm}
\centering
{
\small
\begin{tabular}{|l|C{0.85cm}|C{0.85cm}|C{0.85cm}|C{0.85cm}|}
\toprule
Memory Footprint & \textbf{HT} & \textbf{B+} & \textbf{SA} & \textbf{RX} \\
\midrule
Final size [GB] & 0.68 & 1.23 & 0.54 & 2.78 \\
Overhead during build [GB] & 0 & 1.35 & 0.81 & 4.37 \\
\bottomrule
\end{tabular}
}
\caption{Memory footprint for $2^{26}$~keys.}
\label{tab:build_size}
\vspace*{-0.3cm}
\end{table}

With the build time in mind, let us also inspect the memory footprint of all methods when indexing $2^{26}$~keys.
We differentiate between the space required during construction and the space required afterwards.
From the results in Table~\ref{tab:build_size}, we can see that \textbf{RX} consumes considerably more space during construction than the traditional methods.
After construction, the footprint shrinks noticeably, but is still around twice as high as for \textbf{B+}.
This is a consequence of \textbf{RX} representing each key as a triangle, where other indexes can store each key as-is.
\textbf{SA}~consumes more space than \textbf{HT} during construction (caused by the out-of-place radix sort), but has zero structural overhead afterwards.
\textbf{HT} consumes slightly more due to a 25\% over-allocation to achieve its target \mbox{load factor}.

\vspace*{-0.1cm}
\subsection{Varying the Key Multiplicity}
\label{ssec:key_multiplicity}
\vspace*{-0.1cm}

\begin{figure}[h]
\includegraphics[width=.92\linewidth]{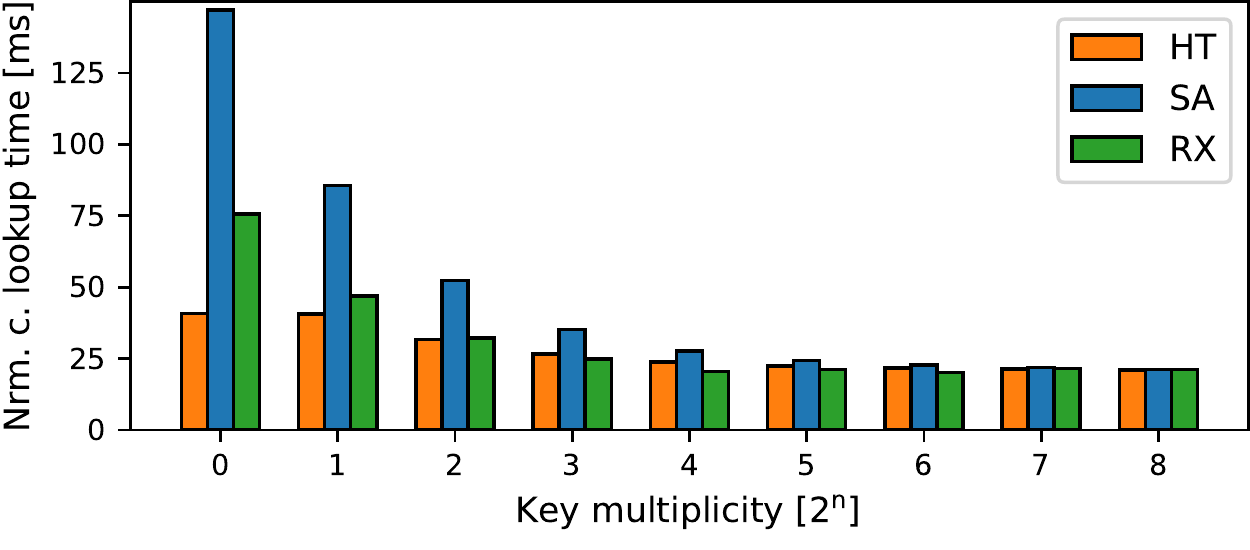}
\caption{Impact of key multiplicity on point lookups.}
\label{fig:key_multiplicity_normalized}
\vspace*{-0.3cm}
\end{figure}

So far, our key set was composed of unique keys only.
In the following, we will introduce duplicates by varying the the key multiplicity and see the impact on the lookup time.
In Figure~\ref{fig:key_multiplicity_normalized}, we vary the key multiplicity from $2^0$ (unique keys) in logarithmic steps to $2^{8}$ (every key appears $256$~times) while keeping the number of point lookups constant.
As the number of results for a point lookup increases with the number of duplicates per key, we normalize the obtained cumulative lookup time by dividing it by the number of duplicates per key.
Note that we cannot show \textbf{B+} in this experiment, because it does not support key duplicates.

From the results we can see that an increased key multiplicity generally favors all indexes.
For \textbf{RX}, duplicate keys lead to the creation of multiple primitives located at exactly the same coordinates in the scene.
Hence, they do not increase the size or complexity of the BVH in any way, but lead only to more ray intersection tests, which are carried out very efficiently in hardware.
Since each lookup ray hits all duplicate primitives in one go, \textbf{RX} handles high key multiplicities well, and marginally wins the comparison for more than $4$~duplicates per key.
According to the profiler, the number of GPU main-memory loads equalizes across the indexes as the multiplicity increases to $2^7$.
Retrieving the value associated with each key now overshadows the cost of traversal.

\vspace*{-0.1cm}
\subsection{\mbox{Ordering Inserts and Lookups}}
\label{ssec:ordering}
\vspace*{-0.1cm}

Until now, we assumed that the indexed keys are inherently unsorted.
However, in practice, the build keys (and their associated values) could be pre-sorted in the column to index.
Also, if sorting is cheap, sorting a batch of point lookups by their requested key might be beneficial.
We therefore investigate the impact of sorted inserts and/or sorted point lookups over the unsorted alternatives.
However, note that sorting is only possible if a sufficient amount of additional GPU memory is available.

\begin{figure}[!ht]
\vspace*{-0.2cm}
\includegraphics[width=.85\linewidth]{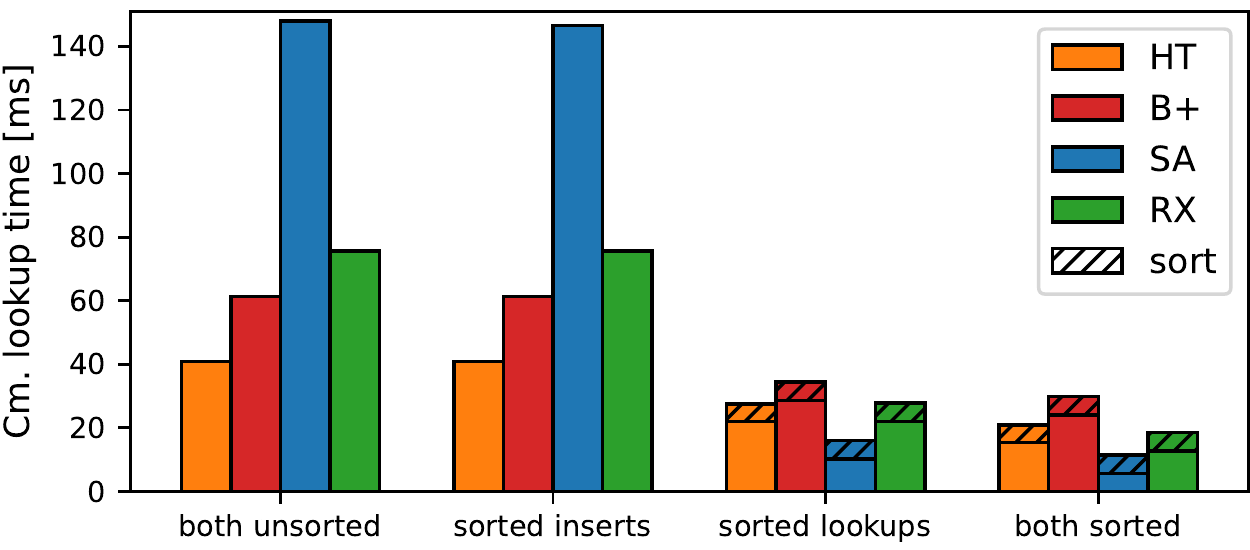}
\caption{Impact of sorted keys and sorted point lookups on lookup performance.}
\vspace*{-0.2cm}
\label{fig:ordering}
\end{figure}

For sorting, we again use CUB's \smalltt{DeviceRadixSort}.
In Figure~\ref{fig:ordering}, we analyze the impact on the cumulative point lookup time for all four combinations. 
When the lookups are unsorted, the build order does not influence lookup times at all.
For the baselines, this is not surprising, since we know that they re-order the keys as part of the build process anyway, either by sorting (\textbf{B+} and \textbf{SA}) or by hashing (\textbf{HT}).
For \textbf{RX}, the same independence is shown by our experiment:
As the run time remains identical, it is very likely that keys are reordered during BVH construction.
In contrast, sorting the point lookups positively impacts all indexes significantly.
This can be attributed to improved memory locality when traversing the index structure, since neighboring lookups are likely to be answered simultaneously by neighboring threads.
As a result, the number of GPU main-memory accesses decreases (between -45\% for \textbf{HT} and -92\% for \textbf{SA}), and just like we explained in Section~\ref{ssec:scaling}, the number of instructions per lookup now limits the throughput.
When both the build set and the lookups are sorted, this locality also extends to the value column:
Neighboring threads look up keys with similar magnitude, which means their associated values will be close together in the sorted build set and even share the same cache line.
Finally, note that GPU-resident sorting is surprisingly cheap in comparison to the actual lookups. 

\vspace*{-0.2cm}
\subsection{Varying the Batch Size for Lookups}
\label{ssec:batching}
\vspace*{-0.1cm}

Next, let us analyze the impact of splitting our $2^{27}$ lookups into multiple batches, which we fire consecutively. 
A submission pattern of multiple smaller batches occurs in practice if (a)~only a limited number of lookups are waiting to be answered simultaneously, (b)~the latency is required be small, or (c)~we want to sort the lookups like in Section~\ref{ssec:scaling}, but there is not enough space available to do it in one go. 

\begin{figure}[!ht]
\vspace*{-0.2cm}
\includegraphics[width=.95\linewidth]{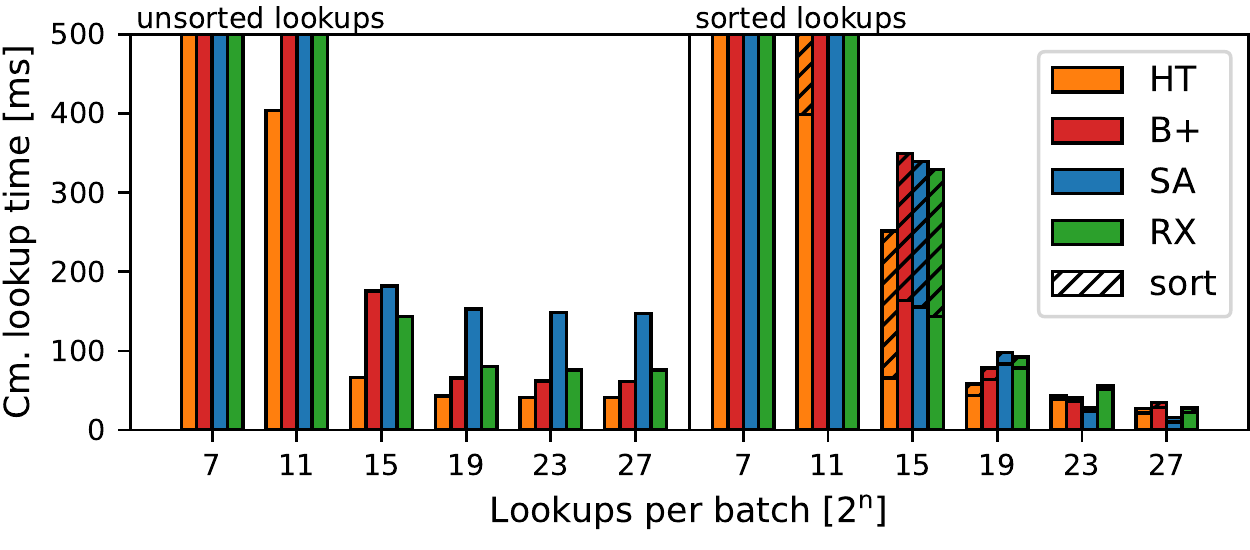}
\caption{Impact of batching lookups (capped at 500 ms).}
\vspace*{-0.2cm}
\label{fig:batching}
\end{figure}

Thus, in Figure~\ref{fig:batching}, we vary the number of batches to submit from $1$ batch ($2^{27}$~lookups per batch) to~$2^{20}$ batches ($128$~lookups per batch) and report the cumulative lookup time on the $y$-axis.
Again, we evaluate both ordered and unordered lookups, to observe the effect of sorting.
We can see that increasing the number of batches generally has a negative effect on all methods.
While the performance for $2^0$, $2^4$, $2^8$, and $2^{12}$~batches remains relatively constant, the performance degrades heavily for more batches due to two reasons:
(1)~From $2^{16}$ batches onwards, each batch is too small to saturate GPU resources (see Section~\ref{ssec:scaling}).
(2)~The total amount of overhead caused by launching CUDA kernels increases, since we have to perform one launch per batch.
Further, we can see that sorting many small batches is more expensive than sorting few larger ones.
In a separate experiment, we confirmed that the runtime of CUB's \smalltt{DeviceRadixSort} stabilizes at a lower bound for batch sizes below $2^{20}$. Consequently, sorting lookups does not improve performance for batch sizes smaller than $2^{19}$.
For \textbf{RX}, a sweet spot is reached for $2^{12}$ batches ($32,768$~lookups per batch), where it surpasses the performance of the other order-based indexes \textbf{B+} and~\textbf{SA}.

\vspace*{-0.1cm}
\subsection{Varying the Hit Rate of Lookups}
\label{ssec:hitrate}
\vspace*{-0.1cm}

\begin{figure}[!ht]
\vspace*{-0.1cm}
\includegraphics[width=.95\linewidth]{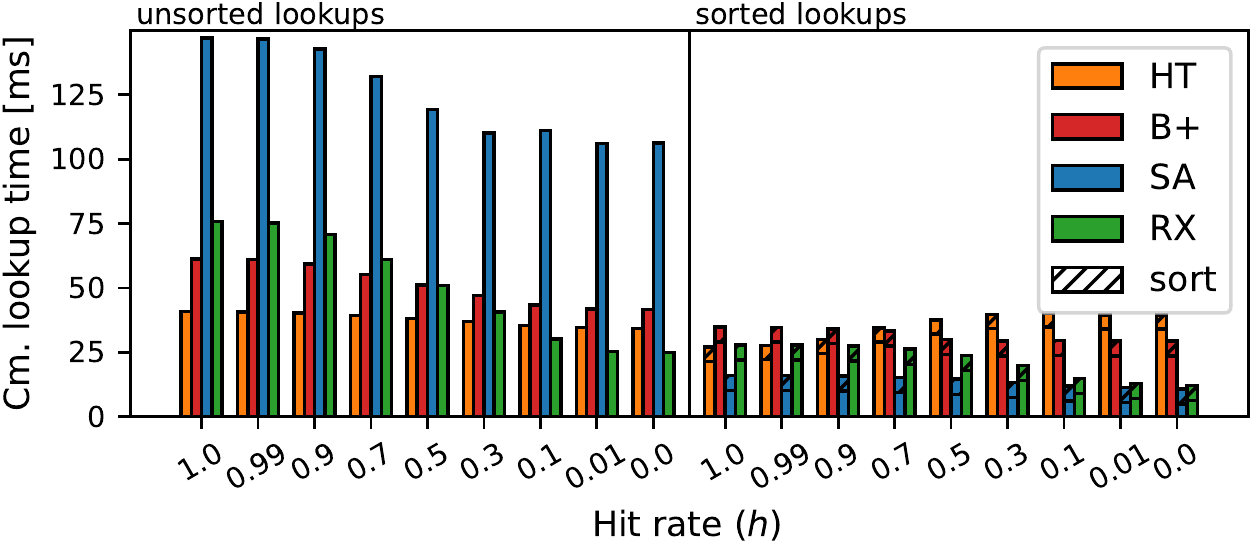}
\caption{Varying the hit rate~$h$.}
\vspace*{-0.2cm}
\label{fig:hitsandmisses}
\end{figure}

Up to this point, we ensured that all lookups were hits and therefore returned a non-empty result.
However, depending on the workload, misses might occur as well, potentially affecting the performance. 
Thus, in Figure~\ref{fig:hitsandmisses}, we vary the hit rate~$h$, i.e., the fraction of lookups that return a non-empty result, on the $x$-axis, and observe the impact on the cumulative lookup time on the $y$-axis.
Again, we test both unordered and ordered lookups.

With a decrease of~$h$, for unsorted lookups, we observe a notable decrease in lookup time for all indexes except for \textbf{HT}.
A constant portion of the decrease can be attributed to the fact that only hits require us to retrieve a corresponding value from the projected column, whereas each miss can skip this step. 
Apart from that, we see a drastically different impact of the hit rate on the different indexes. 
Most notably, \textbf{RX} runs disproportionately faster (up to $3\times$ when going from $h=1.0$ to $h=0.0$), outperforming \textbf{B+} and \textbf{SA} for $h \leq 0.5$ and even \textbf{HT} for $h \leq  0.1$ under unordered lookups.
Here we see the advantage of the BVH over regular search trees:
The BVH traversal can be aborted as soon as no bounding volume at the next lower level covers the searched key.
These early aborts can be seen during profiling as a disproportionate reduction in GPU main-memory accesses (-63\% from $h=1.0$ to $h=0.0$).
In contrast, aborting the traversal early is not possible on regular trees, which always have to do a full traversal.
We confirmed that early abort is indeed the reason for the good performance of \textbf{RX} in an additional separate experiment, where we evaluated the extreme case that all misses lie outside the value range of the key column, i.e., each missed key is smaller/greater than the smallest/greatest key in the key set:
In this situation, \textbf{RX} performs even better than shown in Figure~\ref{fig:hitsandmisses}, as the BVH traversal can already be aborted at the root node.
Lastly, as expected, all methods are generally faster under ordered lookups with \textbf{SA} being the dominant method -- however, as stated before, sorting is only an option if additional memory is available.
In contrast, the performance of \textbf{HT} suffers when the hit rate decreases.
\textbf{HT} implements open addressing, where a miss usually causes longer probe sequences than a hit.
For sorted lookups, and excluding the sorting phase, a hit rate of~$h=0.0$ leads to 36\% more instructions being executed than for $h=1.0$, and the amount of GPU main memory accesses doubling.

\vspace*{-0.2cm}
\subsection{Impact of the Size of Keys}
\label{ssec:key_size}
\vspace*{-0.1cm}

So far, we focused on 32-bit keys, as \textbf{B+} does not support larger keys.
Nevertheless, let us now extend the key size to 64 bits to see how the remaining indexes react.  

\begin{figure}[!ht]
\vspace*{-0.3cm}
\centering
\begin{subfigure}[b]{0.45\linewidth}
\includegraphics[width=\linewidth]{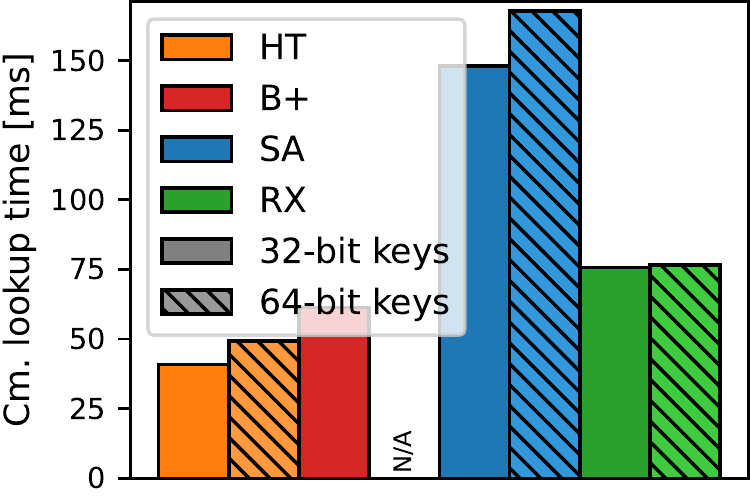}
\caption{Impact on the lookup times.}
\label{fig:key_size:lookup_time}
\end{subfigure}
\begin{subfigure}[b]{0.45\linewidth}
\includegraphics[width=\linewidth]{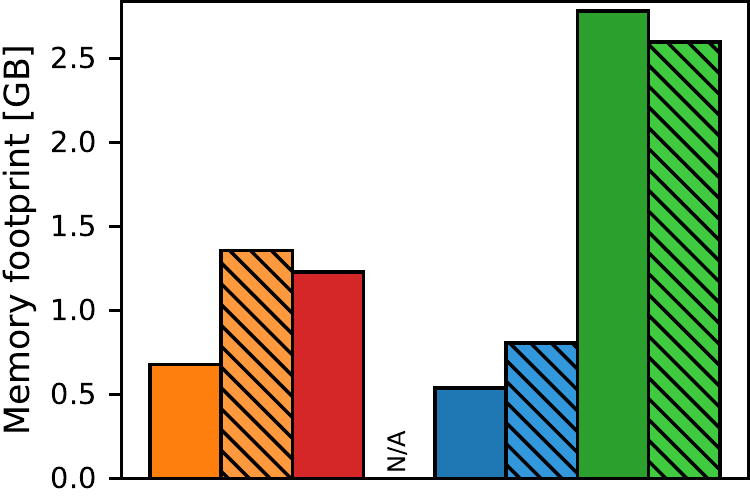}
\caption{Impact on the index size.}
\label{fig:key_size:index_size}
\end{subfigure}
%
\caption{Impact of the key size (32-bit vs 64-bit).}
\vspace*{-0.3cm}
\label{fig:key_size}
\end{figure}

In Figure~\ref{fig:key_size}, we compare the lookup time and index size for 32-bit and 64-bit keys.
We also show the 32-bit results for \textbf{B+} as a point of reference.
Regarding cumulative lookup time, Figure~\ref{fig:key_size:lookup_time} shows that \textbf{RX} is unaffected by the increase in key size, since it does not differentiate between 32-bit keys and 64-bit keys when converting them to triangles.
In contrast, \textbf{SA} and \textbf{HT} slow down under 64-bit keys, which we attribute to the increased cost of 64-bit integer comparisons on GPUs, in combination with the increased memory footprint of the underlying data structure.
Figure~\ref{fig:key_size:index_size} shows the memory footprint of all indexes.
Since \textbf{RX} treats 32-bit keys like 64-bit keys during construction, the size of the BVH is mostly the same, apart from small random variations pertaining to the choice of inserted keys.
In contrast, both \textbf{SA} and \textbf{HT} store each key in its original representation, and therefore, $64$-bit keys lead to a noticeable increase in memory consumption.

\vspace*{-0.2cm}
\subsection{Varying the Skew}
\label{ssec:distribution}
\vspace*{-0.1cm}

We have seen how the indexes perform on a set of uniformly distributed keys, answering uniformly distributed lookups.
In the following, we will introduce skew to both the lookup distribution as well as the key distribution and observe the effects on the indexes.  

\begin{figure}[!ht]
\vspace*{-0.2cm}
\includegraphics[width=.95\linewidth]{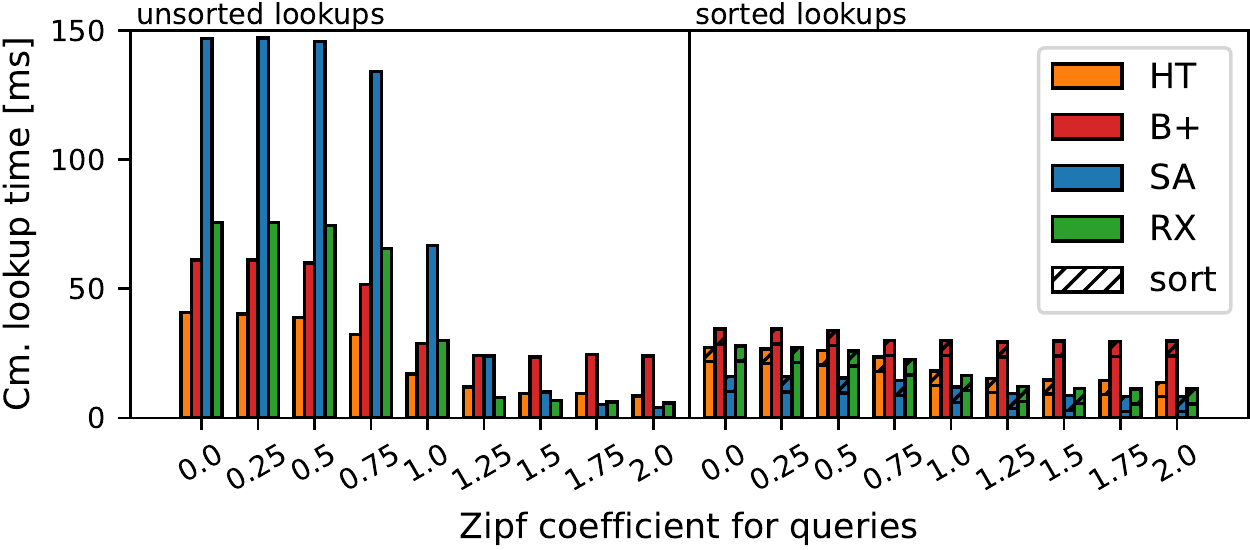}
\caption{Varying the skew of point lookups on uniformly distributed keys.}
\vspace*{-0.3cm}
\label{fig:queryskew}
\end{figure}

In Figure~\ref{fig:queryskew}, we introduce skew to the lookups while keeping the indexed key set uniformly distributed.
The keys of our lookups follow a Zipf distribution, where we vary the Zipf coefficient from~$0.0$ (resembling a uniform distribution) up to a very high skew of~$2.0$ on the $x$-axis and observe the cumulative lookup time on the $y$-axis. Again, we evaluate both sorted and unsorted lookups.

From the results we can see that skewed lookups have a positive effect on the performance of all methods, which holds especially for unsorted lookups.
We attribute this to improved access locality under skew, as the lookups now focus on a smaller key range.
Again, the resulting decrease in memory accesses heavily benefits \textbf{RX} (see Section~\ref{ssec:scaling}),
such that \textbf{RX} outperforms the competing indexes under high lookup skew.
Let us analyze this in detail by comparing \textbf{RX} and \textbf{B+}.
When inspecting the cache hit rate for both methods while varying the skewness in Table~\ref{tab:skew_profiling}, we can see that as long as the skewness is low, the cache hit rate is low as well, and both methods are bandwidth bound.
In this case, \textbf{B+} is superior since it has to read less data from GPU main memory.
However, as soon as the cache hit rate increases due to a higher skewness, both variants become compute bound.
Now \textbf{RX} outperforms \textbf{B+} since it executes around $56\times$ fewer instructions due to the hardware acceleration.
With ordered lookups, the performance improvement is less noticeable, as sorted lookups already ensure good access locality (see Section~\ref{ssec:ordering}).
In a separate experiment, we introduced skew to the key distribution while retaining a uniform lookup distribution (only hits).
However, all methods essentially remained unaffected by a skewed key set.
This is reasonable because each index is able to partition the key set equally well no matter whether the keys are widely spread across the domain or not. 

\begin{table}[h]
    \centering
    \small
    \begin{tabular}{|c|c|c|c|c|c|c|}
    \toprule
    Zipf & \multicolumn{2}{c|}{Hit rate L1/L2 [\%]} & \multicolumn{2}{c|}{Memory read [GB]} & \multicolumn{2}{c|}{Instructions} \\
    coefficient & RX & B+ & RX & B+ & RX & B+ \\
    \midrule
    0.0  & \red{26}/\yellow{44} & \yellow{39}/\yellow{55} & 54.31 & 41.32 & \green{390M} & \red{22B} \\
    0.5  & \red{26}/\yellow{44} & \yellow{39}/\yellow{55} & 53.82 & 40.80 & \green{390M} & \red{22B} \\
    1.0  & \yellow{36}/\green{71} & \green{89}/\green{78} & 22.19 & 16.18 & \green{390M} & \red{22B} \\
    1.5  & \green{82}/\green{90} & \green{92}/\green{93} &  0.85 &  0.79 & \green{390M} & \red{22B} \\
    \bottomrule
    \end{tabular}
    \caption{Impact of skewness on the data transfers and the number of executed instructions (unordered lookups).}
    \label{tab:skew_profiling}
    \vspace*{-0.2cm}
\end{table}

\vspace*{-0.4cm}
\subsection{Answering Range Lookups}
\label{ssec:rq}
\vspace*{-0.1cm}

\begin{figure}[!ht]
\includegraphics[width=.9\linewidth]{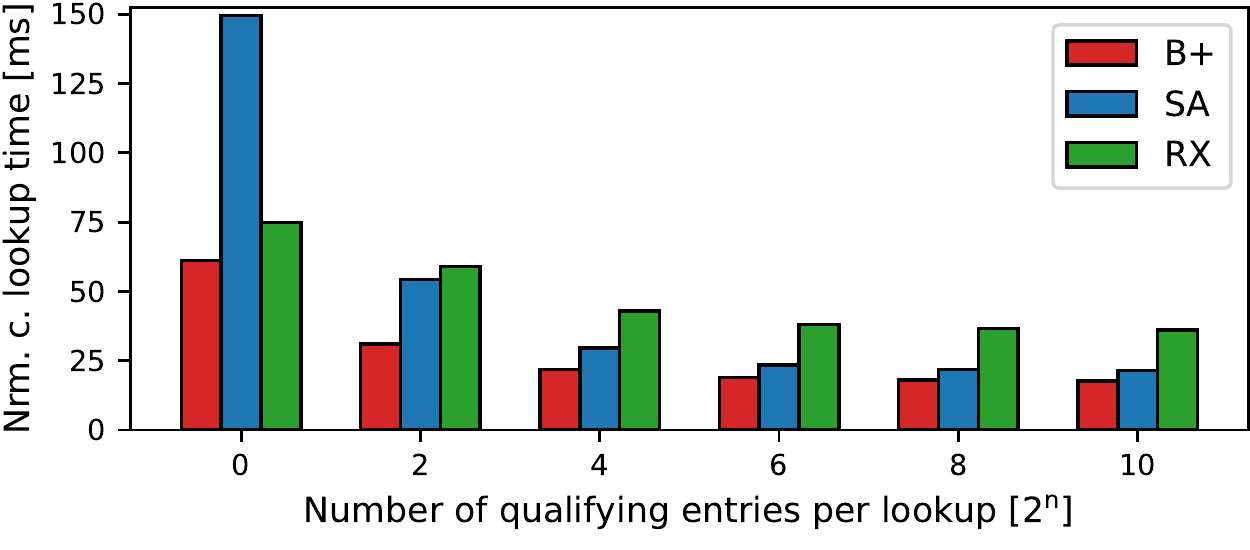}
\caption{Evaluating the cumulative range lookup time.}
\label{fig:range_queries}
\vspace*{-0.2cm}
\end{figure}

So far, we solely evaluated point lookups, where each lookup targeted at most one key, and \textbf{HT} is the fastest index most of the time.
However, as soon as we want to answer range lookups, which are not supported by \textbf{HT}, the order-based indexes \textbf{RX}, \textbf{B+}, and \textbf{SA} can show their strengths.
Thus, in Figure~\ref{fig:range_queries}, we compare these indexes in terms of cumulative range lookup performance while varying the number of qualifying entries from $2^0$ (resembling point lookups) to $2^{10}$.
Note that we normalize the cumulative lookup time by dividing it by the number of qualifying entries per range lookup.

To easily generate range lookups that return a specific number of qualifying entries, we build each index from a column filled with a dense shuffled key set containing \textit{all} integers in $[0, 2^{26}-1]$. 
On such a dense key set, looking up the range $[l^{(i)}, u^{(i)}]$ with a span of $s=u^{(i)} - l^{(i)} + 1$ will return exactly $s$ qualifying entries.
Note that this setup represents the worst-case scenario for range lookups, as all potential keys within each range actually exist.
Thus, it yields an upper bound for the execution time.
From the result, we see that \textbf{B+} yields the best performance across all choices of $s$.
\textbf{RX} initially outperforms \textbf{SA} for small range lookups, but then quickly loses its advantage.
This can be explained by the fact that both \textbf{B+} and \textbf{SA} store the keys in an ordered fashion, and therefore, only have to locate the smallest qualifying key in the index.
All other qualifying keys can be found by traversing the index structure sideways, until the first non-qualifying key is found, where \textbf{B+} implements sideways traversal through a linked list of leaf nodes.
In addition, \textbf{B+} can utilize warp-level aggregation to accelerate the summation of values, giving it an additional advantage over its competitors.
In contrast, \textbf{RX} has to identify each qualifying entry individually by detecting a collision with each triangle that represents a qualifying key. 
Also, we see that the normalized cumulative lookup time of \textbf{RX} decreases when the number of qualifying entries increases. 
This shows that the cost of BVH traversal remains rather constant while varying the number of qualifying entries, whereas the cost for ray intersection tests naturally increases. 

With these results at hand, we can actually approximate the cost of both phases.
Optimistically assuming that exactly one BVH traversal must be carried out per range lookup and that no interleaving takes place, we can create an overdetermined equation system with six equations from the obtained results, where the equation for $2^n$~qualifying entries has the form:
$$\textbf{LookupTime}(2^n) = \textbf{TraversalTime} + 2^n \cdot \textbf{IntersectTime}\text{,}$$
containing the two unknowns \textbf{TraversalTime} and \textbf{IntersectTime}.
Approximating a solution to this equation system using the method of non-negative least squares~\cite{lit:nnls} yields $102.85$ms for \textbf{TraversalTime} and $36.01$ms for \textbf{IntersectTime}, implying that the cost for the BVH traversal dominates the cost for a ray intersection test.

\vspace*{-0.1cm}
\subsection{Varying the Hardware Architecture}
\label{ssec:hardware}
\vspace*{-0.1cm}

Until now, we evaluated all methods on the most recent Ada Lovelace GPU architecture.
By testing the two previous architectures, Ampere and Turing, we can find out how much has changed over the generations, in particular given the varying number of available raytracing cores in different core generations.
Table~\ref{tab:systems} provides an overview of the four test systems spanning three generations of RTX GPUs.
While each system features a different CPU model, remember that all measurements are fully GPU-resident, and therefore, CPU performance can barely influence the results.

\begin{table}[!ht]
\vspace*{-0.1cm}
\centering
{
\small
\begin{tabular}{|l|c|c|c|c|c|}
\toprule
Sys. & GPU & Architecture & VRAM & RTX cores & CPU \\
\midrule
\textbf{S1}  & 4090 & \textit{Ada Lovelace} & 24GB & 128 (3rd gen) & TR 3990X \\
\textbf{S2a} & A6000 & \textit{Ampere} & 48GB & 84 (2nd gen) & i9 12900K \\
\textbf{S2b} & 3090 & \textit{Ampere} & 24GB & 82 (2nd gen) & i7 11700K \\
\textbf{S3}  & 2080Ti & \textit{Turing} & 11GB & 68 (1st gen) & i9 9900K \\
\bottomrule
\end{tabular}
}
\caption{Evaluated GPUs and hardware architectures.}
\label{tab:systems}
\vspace*{-0.3cm}
\end{table}

\begin{figure}[!ht]
\vspace*{-0.2cm}
\includegraphics[width=.95\linewidth]{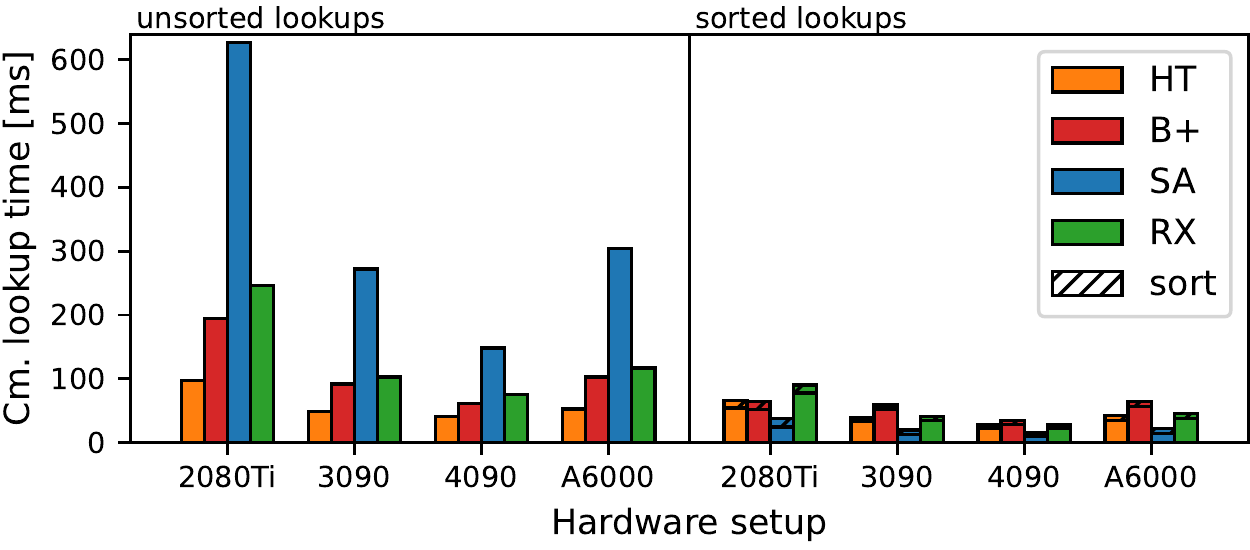}
\caption{Impact of hardware architecture on lookup times.}
\vspace*{-0.3cm}
\label{fig:hardware}
\end{figure}

In Figure~\ref{fig:hardware}, we show the cumulative lookup performance for both sorted and unsorted point lookups on all four test systems. 
We observe a significant performance improvement over the three hardware generations due to an increase in memory bandwidth and number of CUDA cores.
However, we can also see that while \textbf{RX} was not yet competitive on system \textbf{S3}, it is competitive on system~\textbf{S1}.
For sorted lookups, when going from \textbf{S3} to \textbf{S1}, \textbf{RX} shows the most improvement of~$3.23\times$, far more than \textbf{HT} ($2.41\times$), \textbf{SA} ($2.33\times$), and \textbf{B+} ($1.88\times$).
We attribute this difference to a significant increase in the number and performance of the available raytracing cores, which doubled their throughput of ray intersection tests with every generation according to NVIDIA~\cite{lit:ampere_arch, lit:ada_arch}.
For unsorted lookups, this effect is less pronounced, as all methods are limited by memory throughput.
In this case, the improvement for \textbf{RX} of $3.25\times$ is on par with that of \textbf{B+} ($3.17\times$), followed by \textbf{HT} ($2.39\times$). \textbf{SA} improved by~$4.24\times$, but still performs noticeably worse than all other baselines, especially on older hardware.
Since this trend of increasing the number of raytracing cores will likely continue in future generations (such as the upcoming \textit{Grace Hopper} generation), \textbf{RX} might be able to outperform the traditional variants eventually. 

\vspace*{-0.2cm}
\section{Related Work}
\label{sec:related_work}
\vspace*{-0.1cm}

Obviously, NVIDIA RTX was originally designed to accelerate the rendering of complex lighting effects, e.g., transparency, reflection, refraction, and drop shadows, in real-time.
However, apart from this paper, there exist other works outside of rendering that utilized hardware acceleration on RTX-enabled GPUs in creative ways.

For example, one line of work uses raytracing to perform \textit{point containment tests}~\cite{lit:app-containment-1, lit:app-containment-2, lit:app-containment-3, lit:app-containment-4}.
A point containment test determines whether a point lies inside the boundary of a polygon (in 2D) or a polyhedron (in 3D).
One way to implement such a test is by finding the closest intersection with the boundary, which can be solved efficiently on RTX-enabled GPUs.
Another area of application is \textit{time-of-flight imaging}~\cite{lit:app-tof-1, lit:app-tof-2, lit:app-tof-3}.
Time-of-flight sensors compute distances to surrounding objects by measuring the difference between transmission and reception of electromagnetic or acoustic pulses.
Specialized software can then recreate a three-dimensional scene from a sufficient number of measurements.
To test these systems, researchers can simulate the propagation of pulses produced by such a sensor with hardware-accelerated raytracing in a virtual environment.
Also, the problem of performing an efficient \textit{radius search}~\cite{lit:app-neighbor-1, lit:app-neighbor-2} has been accelerated using RTX.
A radius search locates all points within a fixed radius from a specified point.
One can improve this search with a bounding volume hierarchy to quickly exclude far-away points, which can be delegated to the RTX cores.
Further areas of application include physical simulations for \textit{graph rendering}~\cite{lit:app-graph} and \textit{particle movement}~\cite{lit:app-physics-1, lit:app-physics-2, lit:app-physics-3}.

Note that raytracing cores are not limited to consumer and workstation GPUs, but also integrated into some data-center GPUs, including the NVIDIA T4, A2, A10, A16, A40, and L40, some of which can be accessed easily through various cloud providers.
Future data-center GPUs might even incorporate more specialized versions of these cores, if data management applications start exploiting them.

Independent from NVIDIA RTX, in recent years, there is a trend of turning traditional index structures to \textit{GPU-resident index structures}. 
Typically, this implies optimizations to the memory layout and access patterns, and changes to the memory allocation strategy.
Occasionally, a GPU index requires profound algorithmic changes to alleviate contention issues caused by the enormous number of active threads.
Note that some existing GPU indexes only allow lookups on the GPU, while construction and updates can only be done via the CPU.
GPU hashtables~\cite{lit:hash-warpdrive, lit:hash-warpcore, lit:hash-slabhash, lit:hash-megakv, lit:hash-cudpp1, lit:hash-cudpp2, lit:hash-dycuckoo} provide key-value mappings which yield top-of-the-line performance for point lookups, but require various degrees of over-allocation to perform efficiently, and cannot answer range lookups.
If one only needs to test membership, Bloom filters~\cite{lit:bloomfilter1, lit:bloomfilter2, lit:hash-warpcore} and quotient filters~\cite{lit:quotientfilter} require less space than hashtables, but membership tests can produce false positives.
Radix trees~\cite{lit:radixtree} and comparison-based trees~\cite{lit:fasttree, lit:lsmtree, lit:btree1, lit:btree2} also manage key-value mappings, but additionally support range lookups.
Trees generally perform worse than hashtables since tree traversal entails multiple cache misses, and some trees demand sophisticated in-GPU memory managers to dynamically allocate new nodes.
While our comparison includes a state-of-the-art comparison-based tree, no code for the radix tree was freely available at the time of writing.
If a column is limited to a small amount of discrete values, a GPU bitmap index~\cite{lit:bitmapindex} can also offer fast point and range lookups, while keeping a minimal memory footprint.
Since we specifically devised \textbf{RX} to support all possible 64-bit values in Section~\ref{sec:choices}, a fair comparison with bitmap indexes is not possible.
Spatial queries require more specialized solutions, such as GPU R-Trees~\cite{lit:rtree1, lit:rtree2} or GPU permutation indexes~\cite{lit:permutationindex}.
Since R-Trees also employ bounding volumes, it would have been compelling to compare a software R-Tree to our hardware-accelerated BVH approach.
Unfortunately, at the time of writing, no R-Tree implementation was openly available.
Finally, learned indexes~\cite{lit:learnedindex} achieve high performance on GPUs, since they heavily depend on linear algebra operations, which have been extensively optimized on GPUs, and some GPUs even offer specialized accelerators for these operations.
Regrettably, the corresponding implementation cannot be found online.
Apart from indexing on GPUs, there is has been interesting work proposing entirely or partially GPU-resident DBMS architectures~\cite{lit:arch1, lit:arch2, lit:arch3}.
Also, other DBMS operations like joins~\cite{lit:joins1, lit:joins2, lit:joins3, lit:joins4, lit:joins5} or grouping and aggregation~\cite{lit:qp1, lit:qp2, lit:qp3} have been migrated successfully to GPUs.   

Finally, apart from OptiX, DirectX~\cite{lit:directx-announcement} and Vulkan~\cite{lit:vulkan-announcement} provide specialized APIs to specifically target hardware-accelerated raytracing, which both support our proposed indexing scheme.

\vspace*{-0.2cm}
\section{Lessons Learned \& Conclusion}
\label{sec:conclusion}
\vspace*{-0.1cm}

We presented \textbf{RX} and showed that database indexing can indeed be expressed as a raytracing problem to utilize the built-in hardware acceleration of RTX GPUs.
We analyzed five design dimensions and empirically evaluated that by splitting 64-bit keys into three parts (3D Mode) and by using these parts as vertex coordinates for a compacted triangle BVH, we achieve a good trade-off between space utilization and lookup performance.
Further, we discovered that under point lookups, \textbf{RX} can compete with traditional comparison-based indexes.
In high-miss and high-skew scenarios, \textbf{RX} even outperforms the traditional indexes, including the hashtable.
\textbf{RX} also works well when lookups are submitted in smaller batches, and when the lookup distribution is skewed.
However, we also identified that \textbf{RX} is currently not competitive with traditional indexes in terms of build time, memory footprint, and support for updates.
Thus, \textbf{RX} should be used in a read-only fashion.
Finally, we have seen that \textbf{RX} improves faster than the baselines over multiple hardware generations.
If this trend continues, \textbf{RX} might be able to outperform the baselines on future RTX generations. 

\textbf{Acknowledgements:} This work is supported by NVIDIA's Academic Hardware Grant, which provided us with the A6000 GPU.


\bibliographystyle{ACM-Reference-Format}
\bibliography{sample}

\end{document}